\documentclass[pre,twocolumn,superscriptaddress,amsmath,amssymb,floatfix]{revtex4-2}
\usepackage{graphicx}
\usepackage{bm}
\usepackage[colorlinks=true,citecolor=blue,linkcolor=blue,urlcolor=blue]{hyperref}

\newcommand{\kt}{\tilde\kappa}
\newcommand{\mt}{\tilde\mu}
\renewcommand{\Re}{\operatorname{Re}}
\renewcommand{\Im}{\operatorname{Im}}

\begin{document}

\title{Inertial swarmalators with finite-range coupling in two dimensions}

\author{Anirban Ray}
\email{anirban.chaos@gmail.com}
\author{Subhajit Mandal}
\affiliation{Department of Physics, Gour Mahavidyalaya, Mangalbari, Malda, West Bengal, India (affiliated to the University of Gour Banga)}
\author{Asesh RoyChowdhury}
\affiliation{Department of Physics, Jadavpur University, Kolkata 700032, West Bengal, India}

\date{\today}

\begin{abstract}
We study swarmalators with inertia and finite-range coupling in two dimensions.
For attractive phase coupling an energy function forbids sustained motion; for repulsive coupling every steady state obeys an exact sum rule.
When position and phase share a mass $m$, a mode is destabilized by inertia if and only if $\Re s>-m(\Im s)^2$, where $s$ is its overdamped eigenvalue; this yields the Hopf boundary of the one-dimensional inertial swarmalator in closed form, with an exceptional point as its large-mass limit.
In two dimensions, uniform phase waves are transversely unstable and break into filaments.
Inertia sets filaments that are stable without it into motion through a supercritical Hopf bifurcation, a two-dimensional analog of the one-dimensional thrashing wave, while a separate, irregularly moving state coexists with the static filament below threshold.
Simulations show a moving filament network carrying an extensive population of $\pm1$ phase defects.
\end{abstract}

\maketitle

\section{Introduction}

Swarmalators are mobile oscillators whose positions and phases are coupled in both directions \cite{OKeeffe2017,SarReview2026,SarGhosh2022}.
They generalize earlier models of mobile oscillators, in which motion affects synchronization but not the reverse \cite{Frasca2008,Uriu2013}, and of chemotactic oscillators, in which the coupling acts both ways \cite{Tanaka2007,Iwasa2010}.
Swarmalator behavior has been identified in a wide range of systems.
In biology, Japanese tree frogs adjust their calling and their spatial arrangement together in choruses \cite{Aihara2014}.
Vinegar eels (\textit{Turbatrix aceti}) synchronize their body undulations in dense swarms at the edge of a drop, where the resulting metachronal waves drive fluid circulation \cite{Quillen2021,Quillen2022}.
Sperm cells self-organize into vortex arrays through hydrodynamic entrainment \cite{Riedel2005}, and spinning starfish embryos assemble into living chiral crystals \cite{Tan2022}.
Physical realizations are equally varied.
Magnetic Janus colloids under a rotating field link the synchronization of their rotations to self-assembly \cite{Yan2012}.
Coupled magnetic domain walls move faster when their precessions synchronize, and the authors of that study interpret them as swarmalators \cite{Hrabec2018}.
Chemical micromotors synchronize spontaneously while they move \cite{Zhou2020}, and synchronization has been found in collectively moving inanimate and living active matter alike \cite{Riedl2023}.
Magnetically actuated microrobot collectives switch between reconfigurable collective states \cite{Gardi2022}.
Feedback-controlled colloids realize tunable swarmalators whose coupling is hydrodynamic and therefore decays with distance \cite{Heuthe2025}.
Small robots and drones have been programmed to sync and swarm, with reciprocal or non-reciprocal couplings \cite{Barcis2020,Ceron2024}.

Two features recur across these platforms.
Their interactions are short-ranged, being hydrodynamic, steric or limited by line of sight.
In macroscopic realizations such as robots, drones and vibrated granular particles, momentum is moreover not negligible \cite{Scholz2018,Lowen2020}.
Most swarmalator theory nevertheless assumes all-to-all coupling and overdamped motion.
Within that setting the original 2D model and its extensions show ring and splintered states \cite{OKeeffe2017,OKeeffe2018ring}, active phase waves \cite{Hong2018}, chiral and non-uniform behavior \cite{Ceron2023}, and the effects of thermal noise \cite{Hong2023noise}, delays \cite{Blum2024} and non-reciprocal coupling \cite{Yu2025nonrecip,Degond2022}; swarmalators have also been studied in three and higher dimensions \cite{Yadav2024,Anwar2025,Lizarraga2024}.
Much of the analytic progress has come from simplified models: the one-dimensional (1D) ring model \cite{OKeeffe2022ring,Yoon2022,OKeeffeHong2022,Hao2023,OKeeffe2025continuum}, its normal-form derivation from chemotactic oscillators \cite{OKeeffeNormalForm2026,Tanaka2007}, and a two-dimensional (2D) model on a periodic plane in which the dynamics separates into coupled Kuramoto models \cite{OKeeffe2024solvable}.

Two physical ingredients have been studied only separately and only in 1D.
The first is inertia.
O'Keeffe recently showed that adding a mass to the 1D ring model leaves the synchronization and desynchronization thresholds unchanged, but turns the static phase wave into an unsteady ``thrashing'' wave through a subcritical Hopf bifurcation \cite{OKeeffe2026inertia}.
This parallels the role of inertia in the Kuramoto model, where it produces hysteresis and discontinuous transitions \cite{Tanaka1997,Olmi2014,Gupta2014}.
In the 1D swarmalator model every overdamped attractor is static, so the unsteadiness there is produced by inertia alone.
The study is confined to one spatial dimension, and it closes by suggesting that the model be explored in two or three dimensions \cite{OKeeffe2026inertia}.
To our knowledge, whether inertia produces analogous unsteady states in higher-dimensional swarmalator models has not been examined: the recent review of the field discusses inertia only for Kuramoto oscillators \cite{SarReview2026}, and the studies of swarmalators in two or more dimensions cited above do not include it.
The second is finite interaction range.
Sar, O'Keeffe and Ghosh tuned the coupling range of the 1D model and found multiple ``sync dots'', higher-winding waves and an unsteady state at short range \cite{Sar2025range}.
In 2D, a finite cutoff has been imposed on the spatial interaction only \cite{Lee2021}, and a short-range repulsion added to otherwise global couplings produces oscillatory states \cite{JimenezMorales2020}; in both cases the phase coupling remains global.
Finite-range rules in both channels have been combined with self-propelled, run-and-tumble disks, where they produce gel, labyrinth and frustrated phase-separated states \cite{Adorjani2024} and modify ratchet transport on asymmetric substrates \cite{Albert2026}; there the motion is driven by self-propulsion, whereas in our model it is generated by the couplings alone.
Distance-independent topological coupling in 2D produces local vortices \cite{Gou2026topological}, and winding numbers have been used to classify transitions in 2D swarmalator systems \cite{Louodop2025}.

\begin{figure*}[tb]
\includegraphics[width=\textwidth]{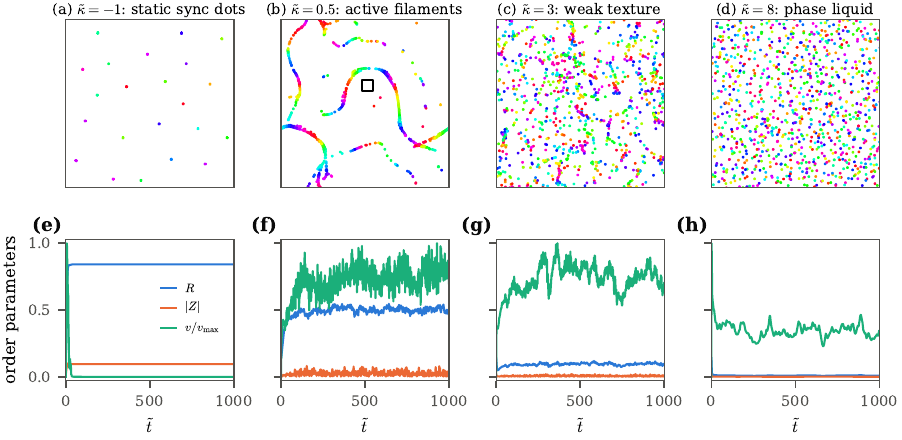}
\caption{Collective states at $\mt=1$, $L=16\ell$, $N=1024$, after $\tilde t=1000$.
(a)--(d) Final snapshots for $\kt=-1$, $0.5$, $3$ and $8$, colored by phase.
Circles and squares mark $+1$ and $-1$ phase defects (Appendix~\ref{app:defects}); none are marked in (a), where the coarse-grained field of isolated clusters is not meaningful.
(e)--(h) Time series of the local phase order $R$, the global order $|Z|=|\langle e^{i\theta}\rangle|$ and the normalized rms speed $v/v_{\max}$ for the same runs, starting from random initial conditions; $v_{\max}=0.32$, $0.46$, $0.20$ and $0.05$ in rescaled units for (e)--(h).}
\label{fig:states}
\end{figure*}

Here we combine both ingredients in 2D.
Our model has three features.
(i) A Gaussian kernel of width $\ell$ makes the interactions local.
(ii) The positional and phase dynamics follow from one pair potential, with mobilities $J'$ and $K'$ that may have either sign.
(iii) Position and phase carry the same mass $m$.
Figure~\ref{fig:states} previews the collective states the model produces: static clusters for attractive phase coupling, and for repulsive coupling a persistently moving network of phase-wave filaments that gives way to a phase-disordered liquid as the repulsion grows.
Feature (ii) is shared by the 1D ring model \cite{OKeeffe2022ring,OKeeffe2026inertia}, so several of our exact results apply to it as well.

This lets us answer, for a two-dimensional model, the question raised by the 1D study.
Inertia does destabilize static phase-wave structures in 2D, and, as in 1D, it does so through an oscillatory mode and with hysteresis.
A straight phase-wave filament that is stable when overdamped is set into persistent motion above a critical mass, first along its own line and then as a meandering filament; this is the 2D counterpart of the thrashing wave.
Three features of the 1D picture do not carry over.
First, the local bifurcation is supercritical rather than subcritical; the hysteresis comes from a separate, irregularly moving state that coexists with the static filament below the threshold.
Second, the 1D static wave itself has no stable 2D analog, because a uniform phase wave in the plane breaks up into filaments for every mass.
Third, with finite-range repulsive phase coupling the overdamped system is already persistently active, so that in the bulk inertia reshapes the unsteady state rather than creating it.
Our model extends the gradient structure of the ring model to the plane.
It is not the solvable 2D model of Ref.~\cite{OKeeffe2024solvable} to which Ref.~\cite{OKeeffe2026inertia} points, and the inertial dynamics of that model remain open.

The paper is organized as follows.
Section~\ref{sec:model} defines the model and its dimensionless parameters.
Section~\ref{sec:exact} derives an energy identity and a general criterion for when inertia alters linear stability, and checks the criterion against the 1D result of Ref.~\cite{OKeeffe2026inertia}.
Section~\ref{sec:linear} analyzes the stationary states of the 2D model: the incoherent state, uniform phase waves and straight phase-wave filaments.
Section~\ref{sec:numerics} presents simulations: the collective states, a numerical regime map, and the scaling of the phase-defect population with system size.
Section~\ref{sec:discussion} summarizes and lists open problems.

\section{Model}\label{sec:model}

We consider $N$ swarmalators with positions $\mathbf{x}_i$ in a periodic square of side $L$ and phases $\theta_i$.
The equations of motion are
\begin{align}
m\ddot{\mathbf{x}}_i+\dot{\mathbf{x}}_i &= -J'\,\frac{\partial H}{\partial \mathbf{x}_i}, \label{eq:x}\\
m\ddot{\theta}_i+\dot{\theta}_i &= -K'\,\frac{\partial H}{\partial \theta_i}, \label{eq:th}
\end{align}
with the pair potential
\begin{equation}
H=-\frac12\sum_{i\neq j} w(|\mathbf{x}_j-\mathbf{x}_i|)\cos(\theta_j-\theta_i),
\qquad
w(r)=\frac{e^{-r^2/2\ell^2}}{n_\ell}.
\label{eq:H}
\end{equation}
Here $n_\ell=2\pi\rho\ell^2$ is the mean number of particles within the kernel and $\rho=N/L^2$.
Written out, Eq.~\eqref{eq:x} reads $m\ddot{\mathbf{x}}_i+\dot{\mathbf{x}}_i=(J'/\ell^2)\sum_j w(r_{ij})(\mathbf{x}_j-\mathbf{x}_i)\cos(\theta_j-\theta_i)$.
For $J'>0$, particles with similar phases attract and those with opposite phases repel.
Equation~\eqref{eq:th} reads $m\ddot\theta_i+\dot\theta_i=K'\sum_j w(r_{ij})\sin(\theta_j-\theta_i)$, a Kuramoto coupling weighted by distance.
It is attractive (synchronizing) for $K'>0$ and repulsive for $K'<0$, as for the contrarian oscillators of Ref.~\cite{HongStrogatz2011}.
We take identical natural frequencies, which a rotating frame sets to zero, and no noise.

The same structure underlies the 1D ring model of Refs.~\cite{OKeeffe2022ring,OKeeffe2026inertia}.
There $H=-(2N)^{-1}\sum_{ij}\cos(x_j-x_i)\cos(\theta_j-\theta_i)$, and the mobilities are the couplings $J'$ and $K'$ of Ref.~\cite{OKeeffe2026inertia}.
The original 2D model of Ref.~\cite{OKeeffe2017} is not of this form: its spatial and phase kernels differ, and it contains a phase-blind attraction and a hard-core repulsion.

We fix $J'>0$.
Measuring lengths in units of $\ell$ and time in units of $\ell^2/J'$ removes $J'$ and $\ell$ from Eqs.~\eqref{eq:x} and \eqref{eq:th}.
What remains are
\begin{equation}
\kt=-\frac{K'\ell^2}{J'},\qquad \mt=\frac{mJ'}{\ell^2},\qquad \frac{\ell}{L},\qquad n_\ell.
\label{eq:groups}
\end{equation}
The factor $\ell^2$ in $\kt$ appears because the spatial force involves a gradient of the kernel while the phase torque does not.
In a local-kernel model, therefore, the bare ratio $K'/J'$ is not dimensionless.
The sign convention makes $\kt>0$ correspond to repulsive phase coupling.
Finally, $n_\ell$ measures how strongly particle discreteness matters.

\section{Exact results}\label{sec:exact}

\subsection{Energy identity, Lyapunov function and sum rule}

Define
\begin{equation}
E=\sum_i\left[\frac{m|\dot{\mathbf{x}}_i|^2}{2J'}+\frac{m\dot\theta_i^2}{2K'}\right]+H .
\label{eq:E}
\end{equation}
Along solutions of Eqs.~\eqref{eq:x} and \eqref{eq:th},
\begin{equation}
\frac{dE}{dt}=-\sum_i\left[\frac{|\dot{\mathbf{x}}_i|^2}{J'}+\frac{\dot\theta_i^2}{K'}\right].
\label{eq:dEdt}
\end{equation}
For $K'>0$, $E$ is bounded below and non-increasing, so it is a Lyapunov function.
By LaSalle's invariance principle every trajectory then approaches the set of stationary points, for any mass.
For $K'<0$ the phase term enters with the opposite sign and unsteady states are allowed.
Averaging Eq.~\eqref{eq:dEdt} over any statistically steady state, in which $\langle dE/dt\rangle=0$, gives the sum rule
\begin{equation}
\frac{\langle|\dot{\mathbf{x}}|^2\rangle}{J'}=\frac{\langle\dot\theta^2\rangle}{|K'|}.
\label{eq:sumrule}
\end{equation}
Positional motion dissipates the energy that the repulsive phase dynamics injects.
Over a finite averaging window, Eq.~\eqref{eq:dEdt} implies that the relative deviation from Eq.~\eqref{eq:sumrule} equals the change of $E$ over the window divided by the energy injected in it, up to time-discretization error.
The deviation therefore measures how far a run is from a steady state.
In the 305 bulk simulations with $K'<0$ reported below, the median deviation is 0.14\%; it is below 1\% in 276 runs and below 2\% in all but 8.
The largest deviations, 3--15\%, occur in slowly evolving overdamped runs at $\kt\le0.25$ (Secs.~\ref{sec:numerics} and \ref{sec:lowk}).

\subsection{When does inertia matter? A criterion}\label{sec:lemma}

Let $\mathbf{q}^*$ be a stationary configuration.
Stack the coordinates as $\mathbf{q}=(\mathbf{x}_1,\dots,\mathbf{x}_N,\theta_1,\dots,\theta_N)$.
Write $D=\partial^2H/\partial\mathbf{q}^2$ for the Hessian at $\mathbf{q}^*$, and $\Gamma=\mathrm{diag}(J'\mathbb{1}_{2N},K'\mathbb{1}_N)$ for the mobility matrix.
Linearizing Eqs.~\eqref{eq:x} and \eqref{eq:th} gives $m\,\delta\ddot{\mathbf{q}}+\delta\dot{\mathbf{q}}=\mathcal{J}\,\delta\mathbf{q}$, with $\mathcal{J}=-\Gamma D$ the Jacobian of the overdamped ($m=0$) system.
Because both channels carry the same mass, the modes of the inertial problem are the eigenvectors of $\mathcal{J}$.
Each eigenvalue $s$ of $\mathcal{J}$ yields two inertial rates $\lambda$ through
\begin{equation}
m\lambda^2+\lambda=s .
\label{eq:quad}
\end{equation}
Setting $\lambda=i\omega$ in Eq.~\eqref{eq:quad} gives the boundary $s=-m\omega^2+i\omega$.
Hence a mode is unstable ($\Re\lambda>0$ for one root) if and only if
\begin{equation}
\Re s>-m\,(\Im s)^2 .
\label{eq:parabola}
\end{equation}
Three consequences follow.

(a) For real $s$, Eq.~\eqref{eq:parabola} reduces to $s>0$, independent of $m$.
Inertia then changes growth rates but not stability thresholds.

(b) For $K'>0$, $\Gamma$ is positive definite and $\mathcal{J}=-\Gamma^{1/2}(\Gamma^{1/2}D\,\Gamma^{1/2})\Gamma^{-1/2}$ is similar to a symmetric matrix.
All $s$ are therefore real, and inertia never moves a stability boundary.
For $K'<0$, $\Gamma$ is indefinite and $\mathcal{J}$ is self-adjoint only with respect to the indefinite metric $\Gamma^{-1}$ ($\mathcal{J}^{\mathsf T}\Gamma^{-1}=\Gamma^{-1}\mathcal{J}$).
In other words, $\mathcal{J}$ is pseudo-Hermitian \cite{Mostafazadeh2002,Bender1998}.
Its eigenvalues are real or come in complex-conjugate pairs, and a pair can leave the real axis only where two real eigenvalues coalesce, at an exceptional point.

(c) The region in Eq.~\eqref{eq:parabola} contains the half-plane $\Re s>0$, so inertia can destabilize a stationary state but never make a linearly unstable one linearly stable.
It destabilizes precisely the oscillatory overdamped modes with $-m(\Im s)^2<\Re s<0$.
This statement concerns the linear stability of stationary states; it says nothing about basins of attraction or about the stability of moving attractors.

The same algebra---a gradient-descent/ascent ``game'' with complex Jacobian eigenvalues, destabilized by momentum---appears in the analysis of min--max optimization \cite{Gidel2019}.

\subsection{Application to the one-dimensional inertial model}
\begin{figure*}[tb]
\includegraphics[width=\textwidth]{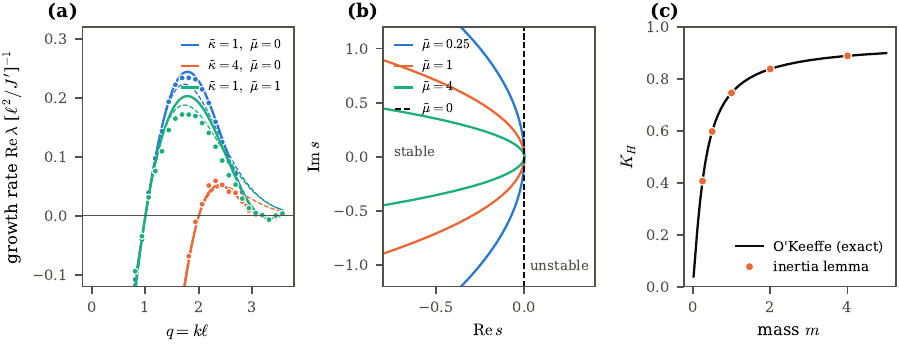}
\caption{(a) Growth rate of the incoherent state: Eq.~\eqref{eq:incoh} with the implemented kernel (solid lines) and with the Gaussian (dashed), and early-time amplitude fits from simulations with $N=16\,384$, averaged over five initial conditions (points; fit windows $\tilde t\le1.5$ for $\mt=0$ and $1\le\tilde t\le3$ for $\mt=1$).
(b) Instability region of a mode with overdamped rate $s$ [Eq.~\eqref{eq:parabola}]: to the right of each parabola.
For $\mt=0$ the boundary is the imaginary axis.
(c) Hopf boundary of the thrashing phase wave in the 1D inertial model: the closed form of Ref.~\cite{OKeeffe2026inertia} (line) and Eq.~\eqref{eq:parabola} applied to the overdamped phase-wave spectrum (points).}
\label{fig:linear}
\end{figure*}

In the sum and difference coordinates $\xi=x+\theta$, $\eta=x-\theta$ of Ref.~\cite{OKeeffe2026inertia}, the overdamped dynamics is $\dot\xi_i=K\,\Im(Ue^{-i\xi_i})+J\,\Im(Ve^{-i\eta_i})$ and $\dot\eta_i=J\,\Im(Ue^{-i\xi_i})+K\,\Im(Ve^{-i\eta_i})$.
Here $U=\langle e^{i\xi}\rangle$, $V=\langle e^{i\eta}\rangle$, and $K=(J'+K')/2$, $J=(J'-K')/2$ are the coefficients of Eqs.~(3)--(4) of that reference.
The static phase wave $\xi_i=0$, $\eta_i=2\pi i/N$ exists for $0<K<J$, which corresponds to $K'<0$.
In Appendix~\ref{app:okeeffe} we diagonalize its overdamped Jacobian exactly.
The spectrum consists of $0$ (translations and phase-wave reparametrizations), $-K$ (all Fourier sectors $|p|\ge2$), and one pair from the $p=\pm1$ sector,
\begin{equation}
s_\pm=-\frac{K}{4}\pm\frac14\sqrt{9K^2-8J^2}.
\label{eq:OKspec}
\end{equation}
The pair is complex for $K<K_{\rm EP}=2\sqrt2\,J/3$ and real for $K_{\rm EP}<K<J$.
At $K_{\rm EP}$ the two eigenvalues and their eigenvectors coalesce: this is an exceptional point.
Inserting Eq.~\eqref{eq:OKspec} into Eq.~\eqref{eq:parabola} gives $4K=m(8J^2-9K^2)$, i.e.
\begin{equation}
K_H(m)=\frac{2\big(\sqrt{1+18m^2J^2}-1\big)}{9m},
\label{eq:KH}
\end{equation}
which is exactly the Hopf boundary of the thrashing wave derived in Ref.~\cite{OKeeffe2026inertia} [Fig.~\ref{fig:linear}(c)].
Equation~\eqref{eq:KH} increases with $m$ and tends to $2\sqrt2J/3$ as $m\to\infty$.
That limit is $K_{\rm EP}$.
The large-mass limit of the inertial instability boundary is therefore the exceptional point of the overdamped spectrum: for $K_{\rm EP}<K<J$ all $s$ are real, and by consequence (a) no mass can destabilize the wave.
This explains why the thrashing region in Ref.~\cite{OKeeffe2026inertia} never reaches the sync threshold $K=J$.
The thresholds of the sync and async states in that model also involve only real $s$ (Eqs.~(27), (28) and (40) of Ref.~\cite{OKeeffe2026inertia}).
Consequence (a) therefore explains why inertia leaves them unchanged.

\section{Stationary states in two dimensions}\label{sec:linear}

From here on we work in the rescaled units of Eq.~\eqref{eq:groups} and treat the density as a continuum.
For a state with uniform density and a phase twist $\theta=\mathbf{Q}\cdot\mathbf{x}$ (including $\mathbf{Q}=0$), the Hessian is diagonal in Bloch wavevector $\mathbf{k}$.
In Appendix~\ref{app:bloch} we show that its $3\times3$ block for displacement $\bm{\xi}$ and phase perturbation $\psi$ is
\begin{equation}
D_{ab}=\Delta\!\left[P_aP_b\,g\right],\quad
D_{a\theta}=-\Delta\!\left[P_a g\right],\quad
D_{\theta\theta}=\Delta[g].
\label{eq:blocks}
\end{equation}
Here $g(\mathbf{P})=e^{-|\mathbf{P}|^2/2}$ is the Fourier transform of the normalized kernel, and $\Delta[f]\equiv f(\mathbf{Q})-\tfrac12\left[f(\mathbf{Q}+\mathbf{k})+f(\mathbf{Q}-\mathbf{k})\right]$ is a second difference.
The overdamped rates are the eigenvalues of $\mathcal{J}=-\mathrm{diag}(1,1,-\kt)\,D$.
Inertia then enters through Eq.~\eqref{eq:quad} with $m\to\mt$.
For the implemented kernel of Eq.~\eqref{eq:wc}, Eq.~\eqref{eq:blocks} holds with $g$ replaced by its Fourier transform $\hat w_c(P)=\int_0^{r_c/\ell}\big(e^{-u^2/2}-e^{-r_c^2/2\ell^2}\big)J_0(Pu)\,u\,du$, because the force is the gradient of the same $w_c$.
The transform $\hat w_c$ equals $0.82$ at $P=0$, exceeds $g$ by up to 13\% near $P\ell=2$, and changes sign at $P\ell=3.17$, beyond which it stays below $2\times10^{-3}$ in magnitude.
We give the results below for the Gaussian, where they take closed form, and state where the implemented kernel changes them.

\subsection{Incoherent state}
The incoherent state has uniform density and uniformly distributed phases.
Linearizing the cold kinetic equation for the first phase harmonic gives
\begin{equation}
\mt\lambda^2+\lambda=\tfrac12\,e^{-q^2/2}\,(q^2-\kt),\qquad q=k\ell .
\label{eq:incoh}
\end{equation}
The right-hand side is real, so by consequence (a) inertia does not move the threshold.
For the implemented kernel, $e^{-q^2/2}$ is replaced by $\hat w_c(q)$; the threshold $q^2=\kt$ is unchanged, the fastest wavenumber shifts by less than 7\% for $0.25\le\kt\le5$, and the peak rate changes by $-8\%$ to $+12\%$.
For $\kt<0$ every wavenumber, including $q\to0$, is unstable, and the system condenses into clusters.
For $\kt>0$ the band $q^2>\kt$ is unstable.
The fastest mode is at $q_*^2=2+\kt$, with overdamped rate $e^{-1-\kt/2}$.
Thus the incoherent state of the Gaussian model is unstable for every $\kt$, but its growth rate falls off exponentially with $\kt$; with the implemented kernel the unstable band is $\kt<q^2<(3.17)^2$, so the statement holds for $\kt<10$, which covers all our simulations.
We measured the initial growth of the phase spectrum in simulations with $N=16\,384$ ($L=64\ell$), averaged over five initial conditions [Fig.~\ref{fig:linear}(a)].
Rates are fitted to the amplitude, i.e.\ half the growth rate of the power, here and throughout the paper.
At $\kt=1$, $\mt=0$ the rates fitted over $\tilde t\le1.5$ agree with the implemented-kernel dispersion to 2\% (median; 6\% at most) across the unstable band, against 6\% (15\% at most) for the Gaussian form.
At $\mt=1$ the growth starts from rest, so we fit over $1\le\tilde t\le3$ and compare with the solution of Eq.~\eqref{eq:incoh} that starts from rest, fitted over the same window; the simulations are then systematically 6\% (median; 13\% at most) below the implemented-kernel prediction.
At $\kt=4$, where the rates are below 0.05, the peak rate is reproduced to 3\% and individual wavenumbers scatter by 10\% (median), against 33\% for the Gaussian form.
In the stable band the measured decay saturates at the particle shot-noise floor.

\subsection{Uniform phase waves break into filaments}
Take a uniform-density wave with $\mathbf{Q}=q\hat{\mathbf{x}}$ and perturb with $\mathbf{k}=k\hat{\mathbf{y}}$.
Then $D_{xy}=D_{y\theta}=0$, and the transverse displacement decouples with
\begin{equation}
s=k^2e^{-(q^2+k^2)/2}>0 .
\label{eq:sheet}
\end{equation}
For the Gaussian kernel this mode is unstable for every $q$, $\kt$ and $\mt$.
Iso-phase lines attract one another, so uniform phase waves break up into filaments.
The fastest transverse wavenumber is $k=\sqrt2/\ell$, well resolved in our boxes; even the longest transverse mode of a box of side $L=24\ell$, $k=2\pi/L$, grows on a time $\tilde t\lesssim70$ for $q\ell\le1.7$.
With the implemented kernel the rate is $k^2\hat w_c(\sqrt{q^2+k^2})$, positive for $q\ell<3.17$; finer waves, with wavelengths below $2\ell$, are only marginally stable, with $|s|<2\times10^{-3}$.

\subsection{Straight filaments}
Next consider a straight line of particles along $\hat{\mathbf{x}}$ with phase $\theta=qx$, normalized so that the line integral of the kernel is unity.
With $G(p)=e^{-p^2/2}$, Eq.~\eqref{eq:blocks} reduces to a transverse branch
\begin{equation}
s_\perp=-\Delta[G],
\end{equation}
and a decoupled longitudinal $2\times2$ block,
\begin{equation}
\mathcal{J}_\parallel=\begin{pmatrix}-\Delta[p^2G] & \Delta[pG]\\ -\kt\,\Delta[pG] & \kt\,\Delta[G]\end{pmatrix}.
\label{eq:Jpar}
\end{equation}
Four instabilities follow.

\emph{Buckling.}
For small $k$, $s_\perp\simeq\tfrac12G''(q)k^2=\tfrac12(q^2-1)G(q)k^2$.
A filament therefore buckles whenever $q\ell>1$, i.e.\ when its phase wavelength is shorter than $2\pi\ell$.

\emph{Long-wave longitudinal modes.}
For small $k$, in units of $\tfrac12k^2G(q)$,
\begin{equation}
\mathrm{tr}\,\mathcal{J}_\parallel=q^4-5q^2+2+\kt(1-q^2),\qquad
\det\mathcal{J}_\parallel=2\kt(1+q^2).
\end{equation}
For $\kt>0$ the determinant is positive, so stability requires a negative trace: $q_-^2<q^2<q_+^2$ with $q_\pm^2=\tfrac12\big[5+\kt\pm(\kt^2+6\kt+17)^{1/2}\big]$.

\emph{Particle-scale phase expulsion.}
As $k\to\infty$ the averages in $\Delta$ vanish, and $\mathcal{J}_\parallel$ then has eigenvalues $0$ and $G(q)(\kt-q^2)$, so an individual particle is expelled from the filament's phase pattern unless $q^2>\kt$.
Combined with the buckling condition, a straight filament can be stable only if $\kt<1$.

\emph{Oscillatory modes.}
At intermediate $k$, $\mathcal{J}_\parallel$ has complex eigenvalue pairs.
These remove most of the remaining window even for $\mt=0$, and by Eq.~\eqref{eq:parabola} inertia removes more of it.

\begin{figure*}[tb]
\includegraphics[width=\textwidth]{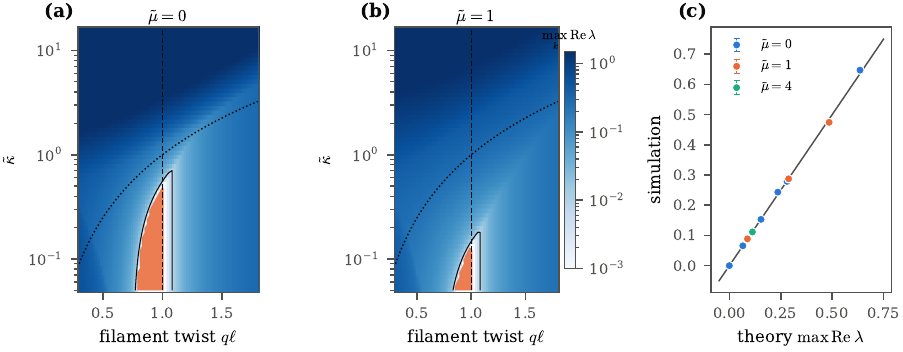}
\caption{Stability of a straight filament with phase twist $q$.
(a),(b) $\max_k\Re\lambda$ from Eq.~\eqref{eq:Jpar} and the transverse branch for $\mt=0$ and $\mt=1$.
Orange: linearly stable.
Dashed line: buckling threshold $q\ell=1$.
Dotted line: particle-scale phase expulsion, $\kt=q^2\ell^2$. Thin solid line: $\max_k\Re\lambda=0.01$, which marks the weakly unstable region hidden by the logarithmic color scale.
(c) Growth rate of the fastest predicted Fourier mode in direct simulations of perturbed filaments ($N=628$--$670$) against the lattice theory at the simulated particle spacing; error bars are standard deviations over three noise realizations.}
\label{fig:filament}
\end{figure*}

\begin{figure*}[tb]
\includegraphics[width=\textwidth]{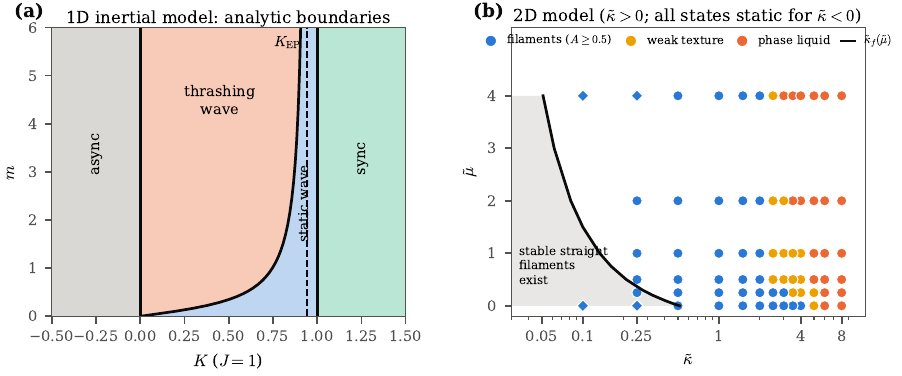}
\caption{State diagram of the 1D model and numerical regime map of the 2D model.
(a) The 1D inertial model of Ref.~\cite{OKeeffe2026inertia} at $J=1$.
The boundaries $K=0$ (async), $K=J$ (sync) and the Hopf curve $K_H(m)$ of Eq.~\eqref{eq:KH} are exact.
As $m\to\infty$ the Hopf curve tends to the exceptional point $K_{\rm EP}=2\sqrt2/3$ (dashed line), beyond which no mass destabilizes the static wave.
(b) The 2D model at $\ell/L=1/24$, $n_\ell\approx25$.
Circles give the simulated regime (three initial conditions each), classified by the texture amplitude of the seed-averaged $C(r)$: filaments ($A\ge0.5$), weak texture ($0.1\le A<0.5$) and phase liquid ($A<0.1$). These thresholds are classification conventions for a smooth crossover (Sec.~\ref{sec:defects}), not phase boundaries; the fit-free peak height of the phase structure factor gives the same map (Sec.~\ref{sec:numerics}). Diamonds are the long low-$\kt$ runs of Sec.~\ref{sec:lowk} ($L=16\ell$).
The solid curve is the analytic boundary $\kt_f(\mt)$ below which an isolated straight filament can be linearly stable (shaded). It is a property of that idealized state, not a boundary of the bulk network, whose filaments are curved, density-modulated and reconnecting. For $\kt<0$ all states are static by Eq.~\eqref{eq:dEdt}.}
\label{fig:regimes}
\end{figure*}

Figure~\ref{fig:filament} shows $\max_k\Re\lambda$ over the plane of $q$ and $\kt$.
Straight filaments are linearly stable only in a small region bounded by $q\ell\le1$, which closes at $\kt_f(\mt)\approx0.51$, $0.30$, $0.13$ and $0.05$ for $\mt=0$, $0.25$, $1$ and $4$; for $\mt\gtrsim2$ the closure point decays roughly as $\kt_f\approx0.2/\mt$ [Fig.~\ref{fig:regimes}(b)].
These results concern an isolated straight filament; in the bulk they identify a mechanism, not a boundary of the regime map.
We checked these predictions with direct simulations of perturbed filaments of 628--670 particles (length $80\ell$, eight particles per $\ell$, initial noise $10^{-8}$, three noise realizations per case), comparing them with the exact lattice version of Eq.~\eqref{eq:Jpar}, which uses the truncated, shifted kernel and the line density of the simulations.
We measure the growth rate of the Fourier mode that the theory predicts to be fastest, with the theory evaluated at the particle spacing $L/N$ of the simulated ring.
In all nine unstable cases it agrees with the predicted rate to within 3.2\% [Fig.~\ref{fig:filament}(c)], including transverse buckling at $\kt=0.25$, $q\ell=1.3$ (0.1\%) and the particle-scale phase expulsion at $\kt=4$ ($+1.8\%$ at $\mt=0$, $-2.2\%$ at $\mt=1$).
Two details matter at $\kt=4$, where the fastest wavelength spans only a few particle spacings.
First, the truncated force is discontinuous at $r_c$, so the theory must use this spacing rather than the nominal $\ell/8$: the two differ by $0.05\%$, but this moves the twentieth neighbor from just outside $r_c$ to $0.999\,r_c$, and with the nominal spacing the predicted rate at $\mt=0$ is $13\%$ lower.
Second, at $\mt=1$ the expulsion modes near $k\ell\approx7$ grow only about 10\% more slowly than the fastest mode ($k\ell\approx3.1$) but start with more power, so they still dominate at the end of the linear regime; their own growth rates agree with the theory at their wavenumbers to within 3\%.
Doubling the line density to sixteen particles per $\ell$ leaves the agreement at $+3\%$ ($\mt=0$) and $-6\%$ ($\mt=1$), and halving $\Delta t$ changes the rates by less than $0.1\%$.
In the one predicted-stable case ($\kt=0.25$, $q\ell=0.95$, $\mt=0$) no growth is seen: the fitted rate of the slowest-decaying mode is $-0.0002\pm0.0010$ (three realizations), against a predicted $-0.0004$.
The same filament becomes unstable at $\mt=1$ and $\mt=4$, with rates $0.088$ and $0.111$ against the predicted $0.088$ and $0.112$.

Together these results show that for $\kt\gtrsim0.5$ no uniform-density state and no straight filament is a stable stationary state, and for $\mt>0$ this holds down to smaller $\kt$.
This is consistent with the persistent dynamics found below.
A stability analysis of curved or interacting filaments remains open.

\section{Simulations}\label{sec:numerics}

We integrate Eqs.~\eqref{eq:x} and \eqref{eq:th} with a semi-implicit Euler scheme that treats the damping implicitly, so that $m=0$ reduces exactly to the overdamped Euler step.
We use $\Delta t=0.05/J'$ for bulk systems and $0.01/J'$ for single filaments.
We repeated nine bulk runs (three parameter points, three initial conditions each) and nine filament runs with half the time step.
In the bulk, $R$, the texture amplitude and wavelength, and the sum rule change by less than 3\%, and the speed, defect density and birth rate change by at most 1.7 standard errors of the seed-to-seed scatter.
For filaments, the linear growth agrees to better than 1\%, onset times are unchanged, and the saturated kinetic energy agrees to within its fluctuations.
The simulations replace $w$ in Eq.~\eqref{eq:H} by the truncated and shifted kernel
\begin{equation}
w_c(r)=\frac{e^{-r^2/2\ell^2}-e^{-r_c^2/2\ell^2}}{n_\ell}\quad(r<r_c),\qquad w_c(r)=0\quad(r\ge r_c),
\label{eq:wc}
\end{equation}
with $r_c=2.5\ell$ unless stated otherwise.
The forces and torques are the exact derivatives of the resulting $H$:
\begin{align*}
m\ddot{\mathbf{x}}_i+\dot{\mathbf{x}}_i&=\frac{J'}{\ell^2}\sum_{j:\,r_{ij}<r_c}\frac{e^{-r_{ij}^2/2\ell^2}}{n_\ell}\,(\mathbf{x}_j-\mathbf{x}_i)\cos(\theta_j-\theta_i),\\
m\ddot\theta_i+\dot\theta_i&=K'\sum_{j}w_c(r_{ij})\sin(\theta_j-\theta_i).
\end{align*}
The constant shift drops out of the force, which therefore carries the unshifted Gaussian, while the torque carries $w_c$.
A finite-difference test of the pair energy computed by the code reproduces its forces and torques to a relative accuracy of $6\times10^{-8}$.
The shift keeps $H$ continuous, which Eq.~\eqref{eq:dEdt} requires; without it the sum rule is violated by $\sim$13\%.
The gradient of $H$, however, remains discontinuous: the force on a pair still jumps to zero at $r_c$.
The shift also lowers the integrated weight of the kernel, $\rho\int w_c\,d^2r=1-(1+r_c^2/2\ell^2)\,e^{-r_c^2/2\ell^2}=0.819$ instead of 1, and changes its Fourier transform (Sec.~\ref{sec:linear}); the filament comparisons below use the implemented kernel exactly.
In disordered states this is immaterial, but in a perfect lattice filament with a neighbor just inside $r_c$ it matters (Sec.~\ref{sec:ignite}).
With the shift, the sum rule holds to within 1\% in 207 of the 216 scan runs with $\kt>0$; the largest deviations, 3--6\%, occur in the slowly evolving overdamped runs at $\kt=0.25$.
Forces are computed with cell lists, in parallel (OpenMP), in a Fortran~90 code.
Unless stated otherwise, $\rho=1$, $\ell=2$ (so $n_\ell\approx25$) and $L=24\ell$ ($\ell/L=1/24$, $N=2304$).
Initial positions and phases are uniformly random; we average over $250\le\tilde t\le500$ and over three initial conditions.
In this window the integrated autocorrelation times of the rms speed and of $R$ are typically $10$ and at most $42$ rescaled time units, so each window spans at least six, and typically 25, correlation times.
Throughout the paper, error bars and $\pm$ values are standard deviations over independent initial conditions or noise realizations, not standard errors; the standard error of a mean over $n$ runs is smaller by $\sqrt n$.
Quantities quoted without an uncertainty come from single runs.
The code and data are available as described under Data availability.

We characterize the states with two diagnostics.
The first is the phase pair correlation $C(r)=\langle\cos(\theta_i-\theta_j)\rangle_{|\mathbf{x}_i-\mathbf{x}_j|=r}$, which we fit to $A\cos(2\pi r/\Lambda)e^{-r/\xi}$ for $r>0.4\ell$, with $\xi\ge\ell/2$, to obtain a texture amplitude $A$ and a wavelength $\Lambda$.
Textured states have $\Lambda\ge2.6\ell$; in the phase liquid the only structure is a weak nearest-neighbor anticorrelation, which the fit captures with $\Lambda\approx1.6$--$2\ell$, and we set $A=0$ whenever $\Lambda<2.3\ell$.
We classify states as filaments ($A\ge0.5$), weak texture ($0.1\le A<0.5$) or phase liquid ($A<0.1$).
These cuts are conventions for a continuous crossover, but the resulting map is robust.
Moving the $\Lambda$ cut anywhere between $2.0\ell$ and $2.6\ell$ changes none of the 72 cells of the regime scan, and moving either $A$ threshold by $0.05$--$0.1$ shifts boundaries by one grid step of $\kt$ (2--8 cells).
As a fit-free check we computed the phase structure factor $S(\mathbf{k})=N^{-1}|\sum_je^{i(\theta_j+\mathbf{k}\cdot\mathbf{x}_j)}|^2$ of the final configuration of one run per cell.
Its peak height $S_*$ separates the three classes without overlap ($S_*=31$--$116$, $4.4$--$26$ and $1.5$--$3.6$), so cuts at $S_*=4$ and $28$ reproduce the map cell for cell.
The second is the number of phase defects, defined as $\pm1$ windings of the coarse-grained field $Z(\mathbf{x})=\sum_jG_s(\mathbf{x}-\mathbf{x}_j)e^{i\theta_j}$.
A winding is counted only when the loop around it lies in a locally ordered and populated region (Appendix~\ref{app:defects}).
As a control, we randomly permute the phases at fixed positions.
This yields no defects in 385 of 389 bulk runs and at most $1.3\times10^{-4}$ per $\ell^2$ in the others, two orders of magnitude below the densities reported below, so the counts do not come from coarse-graining noise.

\subsection{Collective states}
\begin{figure*}[tb]
\includegraphics[width=\textwidth]{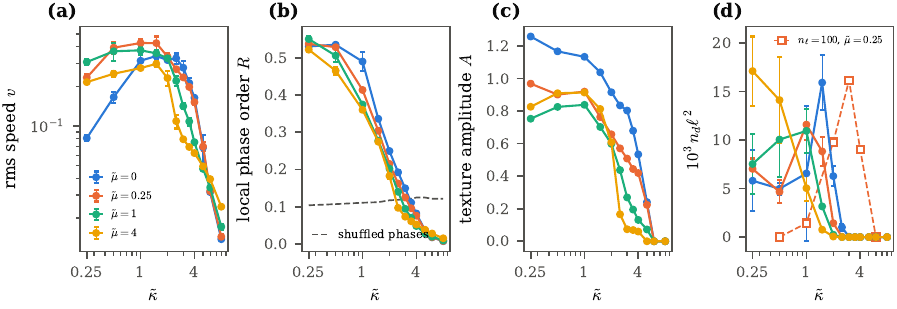}
\caption{Time-averaged order parameters against $\kt$ at $\ell/L=1/24$, $n_\ell\approx25$, for four masses (mean and standard deviation over three initial conditions).
(a) rms speed.
(b) Local phase order $R$; the dashed line is the value for randomly shuffled phases.
(c) Texture amplitude $A$ from fits of $C(r)$.
(d) Defect density. Hollow squares are runs with $n_\ell=100$, $\mt=0.25$ and a coarse-graining scale $\ell/2$, which resolve defects at larger $\kt$ (see text).
For $\kt<0$ (not shown) all states are static, with $R\approx0.84$--$0.90$.}
\label{fig:order}
\end{figure*}

Figure~\ref{fig:states} shows the four collective states together with the time evolution of three order parameters.
These are the local phase order $R=\sum_{\mathbf{x}}|Z(\mathbf{x})|/\sum_{\mathbf{x}}\varrho(\mathbf{x})$, coarse-grained at scale $\ell$; the global Kuramoto order $|Z|=|\langle e^{i\theta}\rangle|$; and the rms speed normalized by its maximum over the run.
The global rainbow order parameters of the long-range models \cite{OKeeffe2024solvable,OKeeffe2026inertia} have no meaning when the coupling is local, which is why we use $R$ instead.
Figure~\ref{fig:order} shows how the time-averaged order parameters vary with $\kt$, and Fig.~\ref{fig:regimes} summarizes the regimes in the $(\kt,\mt)$ plane.

\emph{Static sync dots} ($\kt<0$).
As Eq.~\eqref{eq:dEdt} requires, motion stops: by $\tilde t=500$ the rms speed has decayed below $10^{-12}$ for every mass in the scan.
The particles condense into clusters of nearly equal phase, and because the coupling is local, clusters far apart can hold different phases [Fig.~\ref{fig:states}(a)].
These are the 2D analogs of the sync dots of Ref.~\cite{Sar2025range}.
After a brief transient the speed collapses by $\tilde t\approx40$.
The local order saturates at $R\approx0.84$, while the global order stays at $|Z|\approx0.1$ because distant dots hold different phases [Fig.~\ref{fig:states}(e)].

\begin{figure}[tb]
\includegraphics[width=\columnwidth]{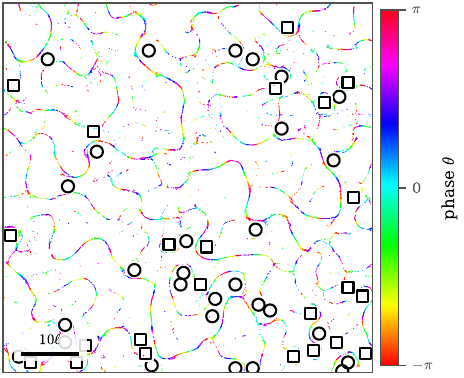}
\caption{Filament network at $\kt=0.5$, $\mt=1$ in a large system ($L=64\ell$, $N=16\,384$), colored by phase.
Circles and squares mark $+1$ and $-1$ phase defects, which sit in voids enclosed by filament loops. The bar is $10\ell$.}
\label{fig:large}
\end{figure}

\emph{Active filaments} ($0<\kt\lesssim2$--$4$).
Particles gather into thin filaments along which the phase winds, like 1D phase waves embedded in the plane [Figs.~\ref{fig:states}(b) and \ref{fig:large}].
The filaments move, bend, break and reconnect without coarsening over the full length of our runs.
The speed fluctuates persistently at about $0.75$ of its maximum.
The local order settles at $R\approx0.50$ while the global order stays near zero ($|Z|\approx0.03$), as expected of a texture whose phase winds along every filament [Fig.~\ref{fig:states}(f)].
The texture scale is set by the linear instability of the incoherent state.
The centroid $k_*$ of the peak of $S(\mathbf{k})$ is $k_*=(0.75$--$1.0)\,q_*$, with $q_*\ell=\sqrt{2+\kt}$ the fastest linear wavenumber of Eq.~\eqref{eq:incoh}, and it is within 6\% of $q_*$ for $\kt\ge2$; at smaller $\kt$ the texture is slightly coarser.
The fitted wavelength is larger than $2\pi/k_*$ by a nearly constant factor of 1.2 (1.09--1.34), because a one-dimensional cosine is fitted to an isotropic two-dimensional correlation.
It decreases from $\Lambda\approx6\ell$ at $\kt=0.25$ to $3.4\ell$ at $\kt=3$ and is almost independent of $\mt$.
The $\pm1$ defects sit in the voids enclosed by filament loops: the phase winds by $2\pi$ around such a loop.

\emph{Weak texture} and \emph{phase liquid} (larger $\kt$).
The texture amplitude $A$ decreases with $\kt$ until $C(r)$ shows no spatial structure beyond the nearest neighbors [Fig.~\ref{fig:states}(c),(d)].
The local order drops accordingly, to $R\approx0.10$ at $\kt=3$ and $R\approx0.01$ at $\kt=8$ [Fig.~\ref{fig:states}(g),(h)].
The latter is below the value $R\approx0.12$ obtained for randomly shuffled phases, so neighboring phases are actively anticorrelated.
The particles still move, slowly: at $\kt=8$ the rms speed is $0.01$--$0.02$ in rescaled units, compared with $\approx0.3$ in the filament state.
This residual motion is a discreteness effect: at $\kt=6$ and $\mt=0.25$ the rms speed falls from $0.057$ at $n_\ell=12.5$ to $0.019$ at $n_\ell=50$, roughly as $n_\ell^{-0.8}$, whereas in the textured states it does not depend on $n_\ell$ (Appendix~\ref{app:nell}).
The anticorrelation persists at every resolution ($R$ stays at 12--20\% of its shuffled value), so in the continuum limit we expect the phase liquid to be a static state with locally anticorrelated phases.

Figure~\ref{fig:order} collects these trends.
The rms speed is largest for $\kt\approx0.5$--$1.5$ and falls by more than an order of magnitude by $\kt=8$ [Fig.~\ref{fig:order}(a)].
The local order $R$ decreases monotonically with $\kt$ and crosses the shuffled-phase baseline between $\kt\approx2.5$ and $3.5$, earlier for larger $\mt$ [Fig.~\ref{fig:order}(b)].
The texture amplitude $A$ and the detected defect density both decrease with $\kt$ [Fig.~\ref{fig:order}(c),(d)], and both vanish at smaller $\kt$ when $\mt$ is larger: $A$ stays above $0.5$ up to $\kt=4$ for $\mt=0$ but only up to $\kt=2$ for $\mt\ge0.5$.

\emph{Role of inertia.}
Inertia does not create the activity.
The overdamped system ($\mt=0$) is already active for $\kt>0$, as Eq.~\eqref{eq:dEdt} permits.
Its effect is quantitative.
At $\kt=0.5$, going from $\mt=0$ to $\mt=1$ raises the rms speed by a factor of 2.2--2.5, the defect density by a factor of 2--3, and the defect birth rate by a factor of 5--10; the ranges span the two time steps of our convergence check, with three initial conditions each.
Inertia also lowers the value of $\kt$ at which the texture decays [Fig.~\ref{fig:regimes}].
The analysis of Sec.~\ref{sec:linear} further predicts that inertia should extend the active regime to lower $\kt$, by removing the stable straight filaments that exist for $\kt<\kt_f(\mt)$.
We test this prediction in Sec.~\ref{sec:lowk}.

\subsection{Inertia and transients}
\begin{figure*}[tb]
\includegraphics[width=\textwidth]{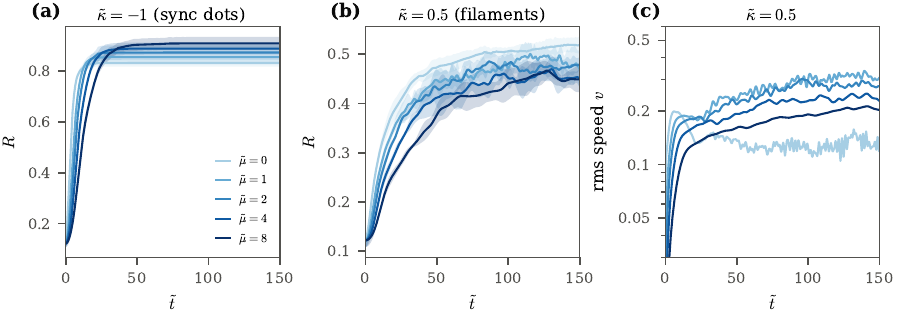}
\caption{Effect of inertia on transients from random initial conditions ($L=16\ell$, $N=1024$), for $\mt=0,1,2,4,8$ (light to dark); means over five initial conditions, with standard deviations shaded in (a) and (b).
(a) Local phase order for $\kt=-1$: inertia slows the approach to the static sync dots.
(b) Local phase order and (c) rms speed for $\kt=0.5$: with inertia the speed keeps growing while the phase texture forms, and the texture is weaker for larger $\mt$.}
\label{fig:transients}
\end{figure*}

Figure~\ref{fig:transients} shows how inertia affects the approach to the asymptotic states from random initial conditions, in the spirit of Fig.~3 of Ref.~\cite{OKeeffe2026inertia}.
For $\kt=-1$ the local order rises monotonically to its final value, and inertia slows the approach.
The time to reach 90\% of the final value grows from $\tilde t\approx7$ at $\mt=0$ to $\tilde t\approx25$ at $\mt=8$.
The final order also increases slightly with $\mt$, from $R=0.83\pm0.02$ to $0.91\pm0.03$ (five initial conditions): heavier particles coalesce into fewer and more compact dots before the motion stops.
For $\kt=0.5$ the speed rises within $\tilde t\approx10$ for every mass, as expected from the mass-independent instability of the incoherent state [Eq.~\eqref{eq:incoh}], although more slowly for larger $\mt$: at $\tilde t=10$ it is $0.19$ for $\mt=0$ and $0.09$ for $\mt=8$.
In the overdamped case the speed then relaxes to a plateau $v\approx0.13$.
With inertia it keeps rising while the phase texture forms, and reaches a plateau 1.6--2.4 times higher only at $\tilde t\approx65$--$75$.
The texture builds up on the same time scale: the local order reaches 90\% of its final value at $\tilde t\approx55$--$85$.
Its final value decreases with mass, from $R=0.52$ at $\mt=0$ to $0.45$ at $\mt=8$, in line with the decrease of the texture amplitude in Fig.~\ref{fig:order}(c).

\subsection{Inertia ignites a linearly stable filament}\label{sec:ignite}
\begin{figure}[tb]
\includegraphics[width=\columnwidth]{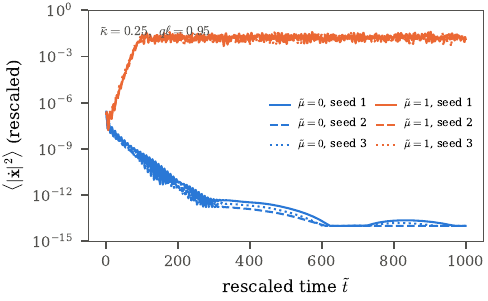}
\caption{Inertia ignites a filament that is linearly stable in the overdamped limit ($\kt=0.25$, $q\ell=0.95$, $N=159$).
Mean squared speed after a perturbation of amplitude $10^{-3}$, for three noise realizations per mass, with kernel cutoff $r_c=2.52\ell$ (see text).
For $\mt=1$ every realization grows at close to the linear rate and settles into persistent motion, first along the line and later as a meandering filament.
For $\mt=0$ every realization relaxes to rest.}
\label{fig:ignite}
\end{figure}

Figure~\ref{fig:ignite} follows the nonlinear fate of a single filament with $\kt=0.25$ and $q\ell=0.95$.
This filament is linearly stable at $\mt=0$ but unstable at $\mt=1$ through an oscillatory mode [Fig.~\ref{fig:filament}].
We perturb it with noise of amplitude $10^{-3}$ and follow three noise realizations for each mass, with the kernel cut off at $r_c=2.52\ell$ instead of $2.5\ell$ for the reason explained below.
At $\mt=1$ the kinetic energy grows in all three close to the predicted rate ($0.081$--$0.084$ in amplitude, against $0.088$) and saturates at $\langle|\dot{\mathbf{x}}|^2\rangle=0.018\pm0.001$ (three realizations).
It continues to fluctuate irregularly by about 30\%, with a dominant period of about 10 rescaled time units.
At $\mt=0$ all three realizations come to rest: the kinetic energy decays, with oscillatory excursions of at most about one decade, below $10^{-13}$ by $\tilde t\approx430$--$530$ and stays there up to $\tilde t=1000$.

The change of cutoff matters for this filament.
In the filament of $N=159$ particles the tenth neighbor lies at $0.998\,r_c$ for $r_c=2.5\ell$, just inside the cutoff, where the radial force jumps.
Displacements of order $10^{-2}\ell$ then carry neighbors back and forth across $r_c$, and the resulting kicks make the particles pair up along the line until the filament breaks into motion.
With $r_c=2.5\ell$, two of three overdamped realizations grow transiently by nearly three decades, to $\langle|\dot{\mathbf{x}}|^2\rangle\sim10^{-5}$ at $\tilde t\approx100$, and escape after $\tilde t\approx440$--$480$ into a weakly moving state with $\langle|\dot{\mathbf{x}}|^2\rangle\approx4\times10^{-4}$--$10^{-3}$.
In paired runs from identical initial conditions this escape occurs only for $r_c=2.5\ell$.
Moving $r_c$ to $2.52\ell$ changes $\mt_c$ by only 1\% and the growth rates by less than 0.1\%, so it removes this artifact without changing the linear problem.
The overdamped filament is therefore stable to small perturbations; the moving state that the escaping runs reach does exist, however, and, as we show next, it coexists with the static filament.
In the inertial case the moving state first forms within the line: up to $\tilde t\approx150$ the filament stays straight to within $10^{-4}\ell$, while its line density becomes strongly modulated (coefficient of variation of the particle count in bins of width $\ell$ about $0.65$, against $0.5$ for random placement).
This in-line state is itself unstable to transverse displacement.
The transverse deviation grows at rate $\approx0.08$ and saturates at an rms value of about $0.5\ell$, with a transverse kinetic energy about 10\% of the longitudinal one, so the long-time state is a meandering but connected filament.
It is a 2D, finite-range counterpart of the thrashing phase wave of Ref.~\cite{OKeeffe2026inertia}, and it arises through the mechanism of Sec.~\ref{sec:lemma}.

\begin{figure*}[tb]
\includegraphics[width=\textwidth]{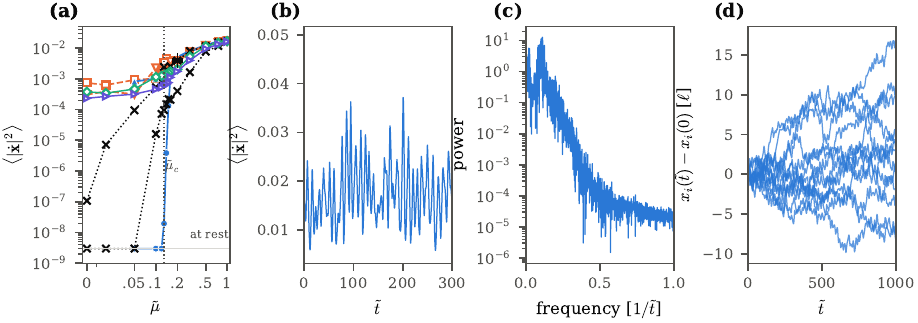}
\caption{The inertia-driven moving filament ($\kt=0.25$, $q\ell=0.95$, $N=159$).
(a) Mean squared speed over the last quarter of each run against $\mt$ (symmetric-logarithmic axis).
Upward sweep from freshly perturbed straight filaments, two noise realizations (blue circles and triangles); downward sweep from the moving state, two realizations (orange open squares and triangles); downward sweeps of longer filaments: $N=317$ (crosses, two realizations), $N=635$ (green diamonds) and $N=1270$ (purple triangles), each the geometric mean of two realizations; upward runs at $\mt=0.2$ for the longer filaments (plus).
Values below $3\times10^{-9}$ (``at rest'') are drawn at that level.
The dotted line marks the linear threshold $\mt_c=0.129$.
(b) Time series (first $300$ of the $1000$ time units analyzed), (c) power spectrum and (d) longitudinal displacement of 13 particles, all at $\mt=1$, after discarding $\tilde t<200$, i.e.\ in the meandering state.}
\label{fig:thrash}
\end{figure*}

To establish the nature of this transition we swept $\mt$ in both directions [Fig.~\ref{fig:thrash}(a)].
The exact lattice theory places the loss of stability of the straight filament at $\mt_c=0.129$.
At $\mt_c$ a complex pair of Bloch modes with wavenumbers $\pm k_c$ ($k_c\ell=2.53$) crosses the imaginary axis.
By Eq.~\eqref{eq:quad} the Hopf frequency is $\tilde\omega_c=\Im\tilde s=0.390$ and the mode crosses with positive speed (Appendix~\ref{app:hopf}).
Because the modes $\pm k_c$ are degenerate, this is a Hopf bifurcation with O(2)-type symmetry, with a traveling-wave and a standing-wave branch \cite{Golubitsky1988}.
Appendix~\ref{app:hopf} reduces the dynamics to its cubic normal form, with coefficients evaluated from exact derivatives of the lattice equations of motion, for filaments of $N=159$ and $317$ particles, and at a second coupling, $\kt=0.15$.
In all three cases both branches are supercritical, and the traveling wave is the stable one; a simulation at $\mt=\mt_c$ reproduces the predicted cubic decay of the traveling-wave amplitude to within 0.1\% [Fig.~\ref{fig:hopf}(b)].
The local bifurcation is therefore supercritical, unlike the thrashing transition of the 1D model.
Its saturation coefficient is small, however: to cubic order $\langle|\dot{\mathbf{x}}|^2\rangle\approx0.25(\mt-\mt_c)$, so the traveling wave would reach the amplitude of the observed moving state within $\Delta\mt\approx0.015$ of threshold, beyond the range of validity of the cubic theory.
For a long filament the corresponding complex Ginzburg--Landau equation has $1+c_1c_3\approx9>0$ and is therefore Benjamin--Feir stable \cite{Aranson2002} (Appendix~\ref{app:hopf}), so the irregularity of the moving state described below is not a modulational instability of the bifurcating wave.
In the upward sweep each run starts from a freshly perturbed straight filament and lasts $\tilde t=2000$; we used two noise realizations.
In the first, the filament stays at rest for $\mt\le0.13$ ($\langle|\dot{\mathbf{x}}|^2\rangle\le2\times10^{-8}$), creeps at $\mt=0.14$, where the linear growth rate is still tiny, and is set into motion for $\mt\ge0.15$.
The second realization reaches the moving state at every mass, down to $\mt=0.05$, through the same cutoff-triggered escape as the overdamped runs with $r_c=2.5\ell$ described above.
With $r_c=2.52\ell$, four upward runs at $\mt=0.05$ and $0.1$ all stay at rest, while the same runs with $r_c=2.5\ell$ all escape after $\tilde t=240$--$850$.
Above threshold both realizations saturate at the same level, which is close to the downward branch: at $\mt=0.3$, for instance, $\langle|\dot{\mathbf{x}}|^2\rangle=(7.5$--$8.2)\times10^{-3}$ upward and $7.5\times10^{-3}$ downward.
Just above threshold the two upward runs with $r_c=2.5\ell$ reach $\langle|\dot{\mathbf{x}}|^2\rangle\approx(1$--$3)\times10^{-3}$ at $\mt=0.16$, against $3.7\times10^{-3}$ on the downward branch; with the steep supercritical growth estimated above, the sweep cannot resolve the traveling wave while its amplitude is small, i.e.\ within $\Delta\mt\lesssim0.01$ of threshold, where its growth is also too slow to saturate within our runs.
In the downward sweep we start from the moving state at $\mt=1$ and lower $\mt$ step by step without resetting the configuration ($\tilde t=800$ per step, two realizations).
The moving state persists below $\mt_c$ and, for this filament, survives all the way to $\mt=0$, where $\langle|\dot{\mathbf{x}}|^2\rangle\approx(3$--$7)\times10^{-4}$.
For every $\mt<\mt_c$ the static filament and a moving state therefore coexist.
This coexistence does not depend on the cutoff coincidence: with $r_c=2.52\ell$ the downward sweep also stays in the moving state down to $\mt=0$, with $\langle|\dot{\mathbf{x}}|^2\rangle\approx2\times10^{-4}$ there, while upward runs at $\mt\le0.1$ stay at rest.

To turn these observations into occupation statistics we repeated the upward runs with $r_c=2.52\ell$ (for which $\mt_c=0.128$) for ten independent noise realizations at each of $\mt=0.10$, $0.12$, $0.14$, $0.16$ and $0.20$, each lasting $\tilde t=2000$.
The outcome is the same for all ten realizations at every mass, so the fraction of runs in the moving state is $0/10$ at $\mt=0.10$, $0.12$ and $0.14$ and $10/10$ at $\mt=0.16$ and $0.20$ (95\% Wilson intervals $[0,0.28]$ and $[0.72,1]$).
Below threshold the perturbation decays monotonically; at $\mt=0.12$, where the decay is slowest ($\langle|\dot{\mathbf{x}}|^2\rangle\propto e^{-0.002\tilde t}$), it falls to $(0.4$--$3)\times10^{-10}$ by $\tilde t=2000$.
At $\mt=0.14$, $0.012$ above threshold, every run is still in the linear growth phase at $\tilde t=2000$ [$\langle|\dot{\mathbf{x}}|^2\rangle=(2$--$10)\times10^{-6}$, growing as $e^{0.0038\tilde t}$], so it is counted as not yet moving.
At $\mt=0.20$ the runs set off after $\tilde t=410$--$480$ and saturate at $\langle|\dot{\mathbf{x}}|^2\rangle=(4.5\pm0.7)\times10^{-3}$ (mean $\pm$ standard deviation over runs), the level of the downward branch [$(3.3$--$4.7)\times10^{-3}$].
At $\mt=0.16$ they set off after $\tilde t=920$--$1110$ but reach only $(3.7\pm0.2)\times10^{-4}$ in $1500<\tilde t<2000$, ten times below the downward branch at the same mass [$(1.5$--$3.7)\times10^{-3}$]; we cannot tell from runs of this length whether this is a distinct low-amplitude state close to threshold or a slow approach to the large-amplitude one.
Thus, in the upward direction, whether a run ends up moving is decided by the sign of the linear growth rate, not by the noise realization.
The downward sweep is equally insensitive to how long each step lasts.
Three further downward sweeps with $r_c=2.52\ell$, one with $\tilde t=800$ per step and two with $\tilde t=1600$ per step, all stay in the moving state down to $\mt=0$, where $\langle|\dot{\mathbf{x}}|^2\rangle=(1.8$--$3.3)\times10^{-4}$, compared with $2.2\times10^{-4}$ for the original sweep with $\tilde t=800$; doubling the dwell time produces no systematic change at any mass.
Because the Hopf bifurcation is supercritical, this moving state is not born at $\mt_c$; it is a separate, large-amplitude attractor, which small perturbations of the static filament do not reach.
The resulting hysteresis resembles that of the subcritical 1D transition \cite{OKeeffe2026inertia}, but its origin is different.

We repeated the sweeps for filaments of 317, 635 and 1270 particles (lengths $80$, $160$ and $320\ell$), with two realizations each.
Upward runs at $\mt=0.2$ saturate at $\langle|\dot{\mathbf{x}}|^2\rangle=(3.4$--$4.5)\times10^{-3}$ for every length, close to the shortest filament.
For $N=635$ and $1270$ the downward sweep reaches $\mt=0$ in the moving state, as for $N=159$.
At $\mt=0$ this state is stationary, with no trend in its kinetic energy over the 800 time units of the step, at $\langle|\dot{\mathbf{x}}|^2\rangle\approx(2$--$5)\times10^{-4}$; the two realizations at $N=1270$ agree to within 5\%.
The filament of 317 particles is an exception: in both realizations its moving state decays to rest at $\mt\lesssim0.05$ (in one of them it is already weaker below $\mt\approx0.5$).
We have not identified the origin of this exception. $N=317$ is the only length whose tenth neighbor lies just outside the cutoff (at $1.001\,r_c$), whereas for $N=159$, $635$ and $1270$ it lies inside by at most 0.2\%, but we have not tested whether this is the cause.
Hysteresis below $\mt_c$ is therefore robust, and in long filaments the moving branch extends to the overdamped limit, where inertia does not create the motion but only removes the static attractor.

Figure~\ref{fig:thrash}(b)--(d) characterizes the long-time, meandering moving state at $\mt=1$.
The mean squared speed fluctuates irregularly by about 30\%.
Its power spectrum is broadband with a maximum near a frequency of $0.1$ in rescaled units (period $\approx10$), in contrast to the nearly periodic thrashing wave of the 1D model.
Individual particles wander along the filament.
Their mean-square displacement is nearly ballistic at short lags (exponent 1.8 for $\Delta\tilde t<10$) and diffusive at long lags (exponent 1.05 for $\Delta\tilde t>30$), with an rms displacement of $2.9\ell$ after $\Delta\tilde t=200$.

\subsection{An extensive population of phase defects}\label{sec:defects}
\begin{figure*}[tb]
\includegraphics[width=\textwidth]{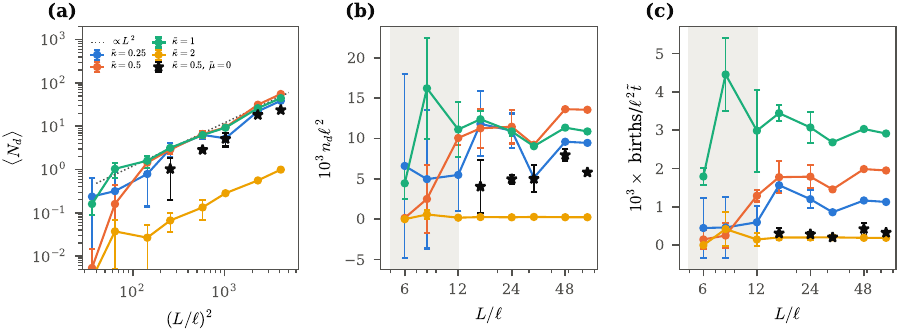}
\caption{Dependence on $\ell/L$ at $\mt=1$, $n_\ell\approx25$.
(a) Mean number of defects against area; dotted line $\propto L^2$.
(b) Defect density.
(c) Defect birth rate.
Error bars are standard deviations over three initial conditions for $L\le24\ell$; larger systems are single runs.
Births are counted between frames $\Delta\tilde t=1$ apart (Appendix~\ref{app:defects}); the $\kt=2$ densities are limited by the resolution of the default detector.
Stars: $\kt=0.5$, $\mt=0$ (three initial conditions for $L\le48\ell$). Shading marks $L<2\Lambda$.}
\label{fig:extensive}
\end{figure*}

Figure~\ref{fig:extensive} shows the dependence on $\ell/L$ at $\mt=1$.
For $\kt\le1$ the mean number of defects grows in proportion to the area once $L\gtrsim12$--$16\ell$, i.e.\ once the box holds about two texture wavelengths ($\Lambda\approx5$--$6\ell$).
At the largest size ($L=64\ell$, $N=16\,384$) it reaches 39, 56 and 45 for $\kt=0.25$, $0.5$ and $1$.
The defect density $n_d\ell^2\approx0.009$--$0.014$ and the defect birth rate, $(0.9$--$3.4)\times10^{-3}$ per $\ell^2$ per unit rescaled time, are then independent of $L$; the one exception, $n_d\ell^2=0.005$ at $\kt=0.25$ and $L=32\ell$, is a single run.
A least-squares fit of $\ln N_d$ against $\ln L$ over all runs with $L\ge16\ell$ gives $N_d\propto L^\alpha$ with $\alpha=1.79$, $2.13$, $1.90$ and $2.05$ for $\kt=0.25$, $0.5$, $1$ and $2$ (95\% bootstrap intervals $[0.99,2.13]$, $[1.75,2.30]$, $[1.54,2.02]$ and $[1.67,2.58]$), consistent with $\alpha=2$; the intervals are wide because the largest sizes are single runs.
For $L\le8\ell$ ($\ell/L\ge1/8$) a system holds about one defect or fewer; this is the crossover toward the long-range limit studied in Refs.~\cite{OKeeffe2024solvable,OKeeffe2026inertia}.
At $\kt=2$ the default detector finds almost no defects ($n_d\ell^2\approx2\times10^{-4}$), a resolution limit discussed below.
With a coarse-graining scale tied to the texture wavelength, $s=\Lambda/5$, the density at $\kt=2$ is $n_d\ell^2\approx7\times10^{-3}$ at both $L=24\ell$ and $48\ell$, comparable to that at $\kt=0.5$ (Table~\ref{tab:defects}).
The overdamped system at $\kt=0.5$ (stars; three initial conditions for $L\le48\ell$, one at $64\ell$) also carries an extensive defect population from $L=16\ell$ on, with $n_d\ell^2=(4$--$8)\times10^{-3}$ and no trend with $L$.
This density is 1.7--2.8 times lower than at $\mt=1$, and the birth rate, $(0.2$--$0.4)\times10^{-3}$, is 5--7 times lower.
Inertia therefore raises the density and especially the turnover of defects, but it is not needed for them to exist.
(A single realization at $L=16\ell$ with almost no defects, $n_d\ell^2=4\times10^{-4}$, is an outlier among the three; it also inflates the fitted exponent, $\alpha=2.8$ with interval $[2.0,4.1]$.)

At $\kt\ge2$ and $n_\ell=25$ few defects are detected for $\mt>0$, although the texture amplitude is still large [Fig.~\ref{fig:order}(c),(d)].
We traced this to resolution: the texture wavelength shrinks as $\kt$ grows ($\Lambda\approx3.9\ell$ at $\kt=2$) and approaches the size of the detection loop.
We therefore repeated the scan at $\mt=0.25$ with four times as many particles per kernel ($n_\ell=100$, $\ell=4$, $L=24\ell$, $N=9216$, $\tilde t=250$) and a coarse-graining scale of $\ell/2$.
The sum rule holds in these runs to within 0.4\% (3\% at $\kt=6$).
The texture itself is unchanged by the fourfold increase of $n_\ell$: $\Lambda$ agrees with the $n_\ell=25$ value to within 2\% at every $\kt$, and $A$ to within about $0.1$. A dedicated test at four resolutions, $n_\ell=12.5$, $25$, $50$ and $100$ at fixed $L=24\ell$ and $\mt=0.25$ (Appendix~\ref{app:nell}, Table~\ref{tab:nell}), confirms and sharpens this.
For $\kt\le2$ the rms speed, $R$, $\Lambda$ and $A$ at all four resolutions agree to within the spread between initial conditions (speed within 10\%, $R$ within $0.01$, $\Lambda$ within 4\%, $A$ within $0.1$).
In the weak texture $\Lambda$ and $R$ are equally well converged, but $A$ grows by about $0.1$ from $n_\ell=25$ to $50$ and then stays constant, so the scan at $n_\ell=25$ slightly underestimates the texture amplitude at $\kt=3$--$4$ and places the texture edge at a slightly smaller $\kt$; $n_\ell=12.5$ is too coarse at $\kt=4$, where it gives $A=0.17$ against $0.48$--$0.51$.
The finer detector finds $n_d\ell^2=0.010$, $0.016$ and $0.009$ at $\kt=2$, $3$ and $4$, against zero in the shuffled control, and none at $\kt=6$, where there is no texture.
Conversely, it finds almost no defects at $\kt\le1$ ($n_d\ell^2\le0.0014$), where the coarser detector finds $0.005$--$0.012$: there $\Lambda\approx5$--$6\ell$, and the small loops fall inside the empty voids enclosed by filaments, where they are masked.
Detected defect densities therefore depend on the ratio of the detection scale to $\Lambda$, and only their scaling with $L$ at a fixed detector, together with their absence where the texture vanishes, is robust.
A coarse-graining scale proportional to the texture wavelength removes the dependence on $\kt$: with $s=\Lambda/5$, defects are found throughout the region with $A\gtrsim0.4$ with a clean shuffled control, and not above the level of the control in the phase liquid (Appendix~\ref{app:defects}).
Taken together, the two detectors show that the defect population is present over the whole region where $A\gtrsim0.5$ [Figs.~\ref{fig:order} and \ref{fig:regimes}(b)], and it disappears with the texture.
To test for finite-size effects we repeated the $\mt=1$ cut at $L=48\ell$ ($N=9216$, one initial condition per $\kt$).
At every $\kt$ from $0.25$ to $8$, $A$ changes by at most $0.04$ and $\Lambda$ by at most 3\% relative to $L=24\ell$, and the decay of $A$ with $\kt$ does not sharpen.
With $s=\Lambda/5$, defects are found at $n_d\ell^2=(6$--$14)\times10^{-3}$ for $\kt\le2.5$ with a clean shuffled control; in the weak texture ($\kt=3$--$4$) the density falls to $(2$--$4)\times10^{-3}$ while the control rises to $(0.3$--$0.7)\times10^{-3}$, and at $\kt=5$ ($A=0.07$) the two are indistinguishable.
At these sizes the disappearance of texture and defects is therefore a smooth crossover; whether it sharpens into a transition in much larger systems we cannot say.

\subsection{Low $\kt$}\label{sec:lowk}
For $\kt<\kt_f(\mt)$ some straight filaments are linearly stable, which raises the question of whether the overdamped system freezes at small $\kt$.
We ran $L=16\ell$ systems from random initial conditions up to $\tilde t=2500$ ($J't=10^4$) at $\kt=0.1$ and $0.25$, with $\mt=0$ and $\mt=4$.
None of them froze.
In each run the mean squared speed in the second half of the averaging window ($1250\le\tilde t\le2500$) was at least as large as in the first half.
The stable filaments of Sec.~\ref{sec:linear} require a specific twist and isolation, and random initial conditions evidently do not reach them on these time scales. This is consistent with the coexistence of static and moving filaments found above for $\mt<\mt_c$.
The run at $\kt=0.1$, $\mt=0$ is still slowly and intermittently evolving at $\tilde t=2500$: its rms speed is only $0.02$, and its sum rule is satisfied only to 15\%.
We therefore continued this case in a separate run to $\tilde t=10^4$, averaging over $5000\le\tilde t\le10^4$.
There it is stationary: the mean squared speed in the two halves of the window agrees to within 0.7\%, the energy per particle drifts by $1\times10^{-6}$ over the window, and the sum rule holds to $0.02\%$.
The system keeps moving, with an rms speed of $0.013$ and local phase order $R=0.54$, and it carries no detected defects.
Inertia nevertheless changes the activity substantially.
Going from $\mt=0$ to $\mt=4$ raises the rms speed by factors of $8.7$ at $\kt=0.1$ and $3.1$ at $\kt=0.25$.
At $\kt=0.1$ the overdamped run has no detected defects, whereas the inertial one has $n_d\ell^2=0.012$ and a birth rate of $1.2\times10^{-3}$; at $\kt=0.25$ the defect density doubles, to $0.015$, and the birth rate rises thirtyfold.
The local phase order is essentially unchanged ($R=0.53$--$0.56$).
Whether the overdamped system eventually freezes at very small $\kt$ remains open.

\section{Discussion}\label{sec:discussion}

We have introduced a two-dimensional swarmalator model with inertia and finite-range coupling that keeps the gradient structure of the one-dimensional ring model.
That structure yields three exact statements.
For attractive phase coupling an energy function decreases monotonically, so all states are static.
For repulsive phase coupling every steady state obeys the sum rule~\eqref{eq:sumrule}.
When position and phase carry a common mass, inertia changes linear stability only through oscillatory overdamped modes [Eq.~\eqref{eq:parabola}], and these exist only in the pseudo-Hermitian regime $K'<0$, where they are born at exceptional points.
In two dimensions, every uniform phase wave breaks up into filaments, and straight filaments are unstable in nearly all of parameter space, with inertia enlarging the unstable region.
A filament that is stable when overdamped loses stability to motion along its line at a critical mass, through a supercritical Hopf bifurcation; at large amplitude the in-line motion is itself unstable, and the filament meanders. A separate, irregularly moving state coexists with the static filament below that threshold, and for all but one of the filament lengths studied it does so down to the overdamped limit, so that there the mass selects which attractor is reached rather than creating the motion.
The generic state is a moving filament network that carries an extensive population of $\pm1$ phase defects.

These results answer, for a finite-range model in two dimensions, whether the inertial unsteadiness of the 1D model \cite{OKeeffe2026inertia} survives beyond one dimension.
Three features carry over.
The relevant instability is oscillatory, and inertia moves its threshold as Eq.~\eqref{eq:parabola} prescribes.
The transition is hysteretic, with static and moving states coexisting below the linear threshold.
And the unsteady state is motion within the phase-wave structure rather than its destruction.
Four features differ.
The local bifurcation is supercritical rather than subcritical: the hysteresis comes from a separate attractor, not from a branch of unstable cycles that ends in a fold.
The object that inertia destabilizes is a filament, not a uniform wave, because uniform waves break up into filaments for any mass.
The moving filament fluctuates irregularly, with a broadband spectrum and particles that wander diffusively along it, whereas the thrashing wave is periodic with a few harmonics. The weakly nonlinear theory does not account for this irregularity, which is a property of the large-amplitude state, and we have not examined whether the mirror-image pairs and chirality switching of the thrashing wave have a filament counterpart.
Most importantly, the bulk is unsteady for $K'<0$ even without inertia, as the energy identity permits: phase repulsion injects energy that positional friction dissipates, in the ratio fixed by Eq.~\eqref{eq:sumrule}.
In 2D with local coupling it is therefore locality, not inertia, that makes the bulk unsteady, and inertia raises the speed and defect turnover and shifts the decay of the texture.
Our analysis also adds to the 1D result itself: the closed-form Hopf boundary follows from Eq.~\eqref{eq:parabola} applied to the overdamped spectrum, and its large-mass limit is the exceptional point $K_{\rm EP}$ (Sec.~\ref{sec:lemma}).

The reason exceptional points control the effect of inertia is simple.
A complex overdamped eigenvalue $s$ means that a perturbation circulates between the positional and phase degrees of freedom while it decays.
Such circulation is possible only when the two channels have mobilities of opposite sign.
Inertia adds a lag to the circulation, and once the energy stored in the momentum, of order $m(\Im s)^2$, outweighs the decay rate $|\Re s|$, the decaying spiral becomes a growing one.
Two general consequences follow.
First, the states that some finite mass can destabilize are exactly those whose overdamped spectrum is complex.
Their boundary, the locus of exceptional points, is therefore a mass-independent stability boundary that can be computed from the overdamped theory alone.
In the one-dimensional model this boundary is $K_{\rm EP}=2\sqrt2J/3$, which is why the thrashing region saturates below the sync threshold.
Second, inertia can never linearly stabilize a stationary state, so adding mass to a gradient-structured swarmalator system can only enlarge the set of linearly unstable stationary states.
The same algebra governs momentum methods in min--max optimization \cite{Gidel2019}, which suggests that the structure is not specific to swarmalators.

The persistent motion we observe requires neither self-propulsion nor noise.
Persistent motion without noise is known in overdamped swarmalators whose couplings lack a gradient structure, such as the active phase waves of Ref.~\cite{Hong2018}; here the gradient structure is intact, and the motion is sustained by the sign of the phase mobility alone.
Energy is injected by the repulsive phase coupling and dissipated by positional friction, and Eq.~\eqref{eq:sumrule} fixes the ratio of the two kinetic energies in any steady state.
The resulting filament network resembles active turbulence \cite{Dunkel2013}, but it differs from that of active nematics \cite{Giomi2015,Shankar2019} in two respects.
Its defects are windings of an internal phase rather than of an orientation field.
They also sit in voids enclosed by filament loops, rather than at points where an ordered fluid is distorted.
Overdamped dynamics already suffices to produce the network.
Inertia raises its speed and defect turnover severalfold and moves the decay of the texture to smaller $\kt$.

These results bear on experiments through the rescaled mass $\mt=mJ'/\ell^2$.
It is the ratio $\mt=\tau_m/\tau_\ell$ of the momentum relaxation time $\tau_m=m/\gamma$ (with $\gamma$ the friction coefficient) to the time $\tau_\ell=\ell^2/J'$ in which the coupling moves a particle across one kernel width.
For micron-sized colloids $\tau_m\sim10^{-7}$\,s, many orders of magnitude below any coupling time, so colloidal and microswimmer swarmalators \cite{Heuthe2025,Riedel2005,Quillen2021} have $\mt\ll1$ and the overdamped predictions apply to them directly.
Uniform phase waves should be unstable to filamentation, and with repulsive phase coupling the steady state should be a moving filament network.
Colloids with feedback-controlled, tunable coupling \cite{Heuthe2025} could test both the filament instability and the sum rule.
In robotic and vibrated-granular platforms the momentum relaxation time is macroscopic \cite{Scholz2018,Lowen2020}, so $\mt=O(1)$ is plausible whenever the coupling acts on a comparable time scale; we have not estimated $\mt$ for a specific platform. In programmable robots and drones \cite{Barcis2020} the mass can even be set by the controller.
In such systems Eq.~\eqref{eq:parabola} could be tested directly: scanning $m$ at fixed coupling should reveal the onset of unsteadiness predicted by the overdamped spectrum, bounded by the exceptional-point locus.
Magnetic domain walls, whose dynamics carry an effective mass and which synchronize their precession \cite{Hrabec2018}, are a further candidate.

Our study has several limitations.
The simulations are noise-free, and the regime scan uses three initial conditions per parameter point at a single system size.
Absolute defect densities depend on the detection scale relative to the texture wavelength; only their scaling with system size is detector-independent.
In perfect filaments, the jump of the truncated force at $r_c$ can trigger escape from the static state; we confirmed the coexistence itself with $r_c=2.52\ell$, where this does not occur.
The moving filament survives in the overdamped limit at three of the four lengths studied (up to 1270 particles), but not at the fourth, for reasons we have not identified.
Convergence in $n_\ell$ has been tested at a single mass, $\mt=0.25$, and a single size, $L=24\ell$, in runs of $\tilde t=250$; in the weak texture and the phase liquid the kinetic energy at $n_\ell=100$ was still rising by 20--60\% between the two halves of the averaging window.
The overdamped state at $\kt=0.1$ has been followed to stationarity in a single run of $1024$ particles, and the texture edge has been followed only up to $L=48\ell$, with one initial condition at that size.

Several questions remain open.
The decay of the texture with $\kt$, and its shift to smaller $\kt$ with inertia, lie outside the linear analysis.
A self-consistent theory in which particle motion acts as an effective temperature for the phases is a natural candidate, since Eq.~\eqref{eq:sumrule} ties the two.
Whether the texture edge is a genuine non-equilibrium transition, and if so its universality class, requires finite-size scaling at larger $n_\ell$.
A coarse-grained theory of moving filaments and their defects would connect the defect statistics to the parameters in Eq.~\eqref{eq:groups}.
If position and phase carry different masses, or if masses vary across particles, Eq.~\eqref{eq:quad} becomes a genuine quadratic eigenvalue problem and our criterion no longer applies directly.
Masses correlated with other particle properties are a natural next step.
The inertial dynamics of the solvable 2D model \cite{OKeeffe2024solvable}, the extension suggested in Ref.~\cite{OKeeffe2026inertia}, also remain to be studied; Eq.~\eqref{eq:parabola} applies to any model with a gradient structure and a common mass, and would be a natural starting point there.
Couplings that break the gradient structure, such as phase lags, natural frequencies, or the kernels of Ref.~\cite{OKeeffe2017}, make $\mathcal{J}$ genuinely non-reciprocal \cite{Fruchart2021,Yu2025nonrecip,Degond2022} and remove the energy identity; whether exceptional points still organize the effect of inertia there is open.
Finally, thermal or active noise \cite{Hong2023noise} turns the deterministic network into a stochastic steady state, and how such noise competes with the self-generated fluctuations fixed by Eq.~\eqref{eq:sumrule} is unexplored.

\begin{acknowledgments}
The authors thankfully acknowledge the One Nation One Subscription (ONOS) initiative of the Government of India, managed by the INFLIBNET Centre, for providing seamless institutional access to the scholarly e-resources and international journals utilized in this research.
\end{acknowledgments}

\paragraph*{Data availability.} The simulation code (Fortran 90), all input files with their random seeds, the raw simulation output, and the analysis and figure scripts (Python) are deposited at Zenodo, \href{https://doi.org/10.5281/zenodo.23855200}{doi:10.5281/zenodo.23855200}.

\appendix
\section{Bloch Hessian of twisted states}\label{app:bloch}
For a pair potential $V(\mathbf{r},\phi)=-w(r)\cos\phi$ with $\phi=\theta_j-\theta_i$, the second variation of $H$ around a state in which every pair sits at $(\mathbf{r},\mathbf{Q}\cdot\mathbf{r})$ is
\begin{equation*}
\delta^2H=\tfrac14\sum_{i\ne j}(\delta\mathbf{u}_j-\delta\mathbf{u}_i)^{\mathsf T}V''_{ij}(\delta\mathbf{u}_j-\delta\mathbf{u}_i),
\end{equation*}
with $\mathbf{u}=(\mathbf{x},\theta)$.
For a Bloch mode $\delta\mathbf{u}_j=\mathbf{U}e^{i\mathbf{k}\cdot\mathbf{x}_j}$ this gives the per-particle block
\begin{equation*}
D(\mathbf{k})=\rho\!\int d^2r\,V''(\mathbf{r})\big[1-\cos(\mathbf{k}\cdot\mathbf{r})\big].
\end{equation*}
The terms odd in $\mathbf{r}$ cancel because $V_{a\theta}=\partial_aw\,\sin(\mathbf{Q}\cdot\mathbf{r})$ is even.
Using $\rho\int\partial_a\partial_bw\,e^{i\mathbf{P}\cdot\mathbf{r}}=-P_aP_bg(\mathbf{P})$ and $\rho\int\partial_aw\,e^{i\mathbf{P}\cdot\mathbf{r}}=-iP_ag(\mathbf{P})$ yields Eq.~\eqref{eq:blocks}.
For a filament, the integral runs along the line and $\partial_y^2w|_{y=0}=-w$, which gives the transverse entry $\Delta[G]$.

\section{Overdamped spectrum of the 1D phase wave}\label{app:okeeffe}
Set $\xi_i=\delta\xi_i$ and $\eta_i=\eta_i^*+\delta\eta_i$, with $\eta_i^*=2\pi i/N$, so that $U=1$ and $V=0$ at the fixed point.
To first order, $U=1+i\bar\xi$ with $\bar\xi=\langle\delta\xi\rangle$, and $V=i\langle e^{i\eta_j^*}\delta\eta_j\rangle$.
Since $V$ is already of first order,
\begin{align*}
\Im(Ue^{-i\xi_i})&=\bar\xi-\delta\xi_i,\\
\Im(Ve^{-i\eta_i})&=\Big\langle\cos(\eta_j^*-\eta_i^*)\,\delta\eta_j\Big\rangle_j .
\end{align*}
Hence
\begin{align*}
\delta\dot\xi_i&=K(\bar\xi-\delta\xi_i)+J\langle\cos(\eta_j^*-\eta_i^*)\delta\eta_j\rangle_j,\\
\delta\dot\eta_i&=J(\bar\xi-\delta\xi_i)+K\langle\cos(\eta_j^*-\eta_i^*)\delta\eta_j\rangle_j .
\end{align*}
The fixed point is invariant under cyclic relabeling, so the Jacobian is block-diagonal in the discrete Fourier modes $(\delta\xi_j,\delta\eta_j)=(b,a)\,e^{ip\eta_j^*}$, $p=0,\dots,N-1$.
Using $\langle\cos(\eta_j^*-\eta_i^*)e^{ip\eta_j^*}\rangle_j=\tfrac12e^{ip\eta_i^*}$ for $p=\pm1$ (zero otherwise) and $\bar\xi=b\,\delta_{p0}$, the Jacobian splits into three kinds of block.
For $p=0$ we have $\dot b=\dot a=0$, giving two zero modes: a uniform translation of $\xi$ and a uniform shift of $\eta$, which reparametrizes the wave.
For $|p|\ge2$ we have $\dot b=-Kb$ and $\dot a=-Jb$, giving the eigenvalues $-K$ and $0$; the neutral $\eta$ modes leave $V=0$ unchanged, so the wave persists as a degenerate family.
For $p=\pm1$,
\begin{equation*}
\frac{d}{dt}\begin{pmatrix}b\\a\end{pmatrix}=\begin{pmatrix}-K&J/2\\-J&K/2\end{pmatrix}\begin{pmatrix}b\\a\end{pmatrix},
\end{equation*}
whose trace is $-K/2$ and determinant $(J^2-K^2)/2$, and whose eigenvalues are given by Eq.~\eqref{eq:OKspec}.
A positive determinant requires $K<J$; at $K=J$ the wave meets the sync branch.
For $0<K<J$ both roots have negative real part, so the wave is linearly stable transverse to its neutral family when overdamped, in agreement with Ref.~\cite{OKeeffe2026inertia}.
Because position and phase carry the same mass, Eq.~\eqref{eq:quad} applies mode by mode.
Only the $p=\pm1$ pair can be complex, so only these modes can be destabilized by inertia, consistent with the thrashing wave, in which the order parameter $|V|$ (denoted $s$ in Ref.~\cite{OKeeffe2026inertia}) oscillates while $|U|$ stays near one.
Counting sectors, the $2N$ eigenvalues are: $0$ with multiplicity $N-1$ (two from $p=0$ and one from each of the $N-3$ sectors with $|p|\ge2$), $-K$ with multiplicity $N-3$, and $s_+$ and $s_-$, each twice (from $p=\pm1$).
Because of the $N-1$ neutral directions the wave is not asymptotically stable in the ordinary sense.
Two are the symmetries (uniform translation and phase shift).
The other $N-3$ redistribute the $\eta_i$ while keeping $V=0$.
All $N-1$ are tangent to an exact manifold of equilibria: every configuration with equal $\xi_i$ and $V=0$ is stationary, since then $U=e^{i\xi}$ and both right-hand sides vanish.
The static wave is one point on this $(N-1)$-dimensional manifold, and our stability statements refer to the directions transverse to it.
When the transverse eigenvalues have negative real parts the manifold is normally attracting, so a small perturbation relaxes to a nearby static wave of the family rather than to the original one.
We confirmed the full spectrum $\{0,-K,s_\pm\}$ with these multiplicities against a numerically computed Jacobian for $N=120$ and $(J,K)=(1,0.3),(1,0.95),(1.5,0.7),(1.5,1.3),(0.8,0.3)$.
These cases include the real branch beyond $K_{\rm EP}$.

\section{Defect detection}\label{app:defects}
The particle field is deposited on a grid and convolved with a Gaussian of width $s$, with $s=\ell$ unless stated otherwise, giving $Z(\mathbf{x})$ and the density $\varrho(\mathbf{x})$; the grid spacing is $s/2$.
A plaquette with winding number $\pm1$ is counted as a defect only if every site on a square loop of half-width about $1.5s$ around it satisfies $|Z|/\varrho>0.2$ and $\varrho>0.3\bar\varrho$, and the winding along that loop equals the plaquette charge.
A defect is counted as born if no defect of the same sign lay within $3s/4$ of it in the previous frame; frames are $\Delta\tilde t=2$ apart in the regime scan and $\Delta\tilde t=1$ in the size scan.
Permuting the phases among particles at fixed positions gives zero defects in 385 of 389 bulk runs and at most $1.3\times10^{-4}$ per $\ell^2$ in the rest.
For $n_\ell=25$ and $s=\ell$, the detection fails once the texture wavelength drops below about $4\ell$; the $n_\ell=100$ runs with $s=\ell/2$ resolve defects for $2\le\kt\le4$ but miss them for $\kt\le1$, where $\Lambda\gtrsim5\ell$ and the smaller loops fall inside empty voids.

\emph{Sensitivity.}
To test which statements depend on the detector, we reanalyzed configurations with a Python implementation that reproduces the defect lists of the simulation code exactly.
We varied one parameter at a time: the coarse-graining scale $s$, the order threshold (0.2 by default), the density mask (0.3) and the loop size.
Table~\ref{tab:defects} gives time averages over frames $\Delta\tilde t=1$ apart at $\mt=1$, for $L=16$, $24$ and $48\ell$ (one run each; 150 frames over $150\le\tilde t\le300$, and 110 frames over $110\le\tilde t\le220$ for $L=48\ell$).
Three statements hold for every setting.
First, the shuffled control gives at most $4\times10^{-5}$ per $\ell^2$.
Second, the density does not decrease between $L=24\ell$ and $48\ell$, so the population is extensive; at $\kt=0.5$ it varies by up to 50\% between sizes, comparable to the seed-to-seed scatter of the regime scan, and at $\kt=2$ the two settings that resolve defects there drop by about 30\% from $L=16\ell$ to $24\ell$ and then stay flat.
Fitted over the three sizes, $N_d\propto L^\alpha$ gives $\alpha=2.0$--$2.4$ at $\kt=0.5$ for every setting and $\alpha\approx1.7$ at $\kt=2$ for those two settings; with one run per size these exponents are indicative only.
Third, applied to the final configurations of the regime scan, no setting with a clean shuffled control finds defects in the phase liquid; the only exception, $s=\ell/2$, also produces comparable numbers of false positives.
Absolute densities, in contrast, change by factors of 2--5 with the threshold, mask and loop size.
The coarse-graining scale matters most: $s=\ell$ misses the defects at $\kt\gtrsim2$, $s=\ell/2$ produces false positives in the control at $n_\ell=25$, and $s=\Lambda/5$ resolves defects wherever $A\gtrsim0.4$ with a clean control; in the weak texture its control is no longer clean (Sec.~\ref{sec:defects}).
Birth rates also depend on the frame interval, because pairs created and annihilated between frames are missed and fast defects are counted as new.
At $\kt=0.5$ and $L=48\ell$, the default detector gives $(2.4,\,2.0,\,1.6,\,1.1)\times10^{-3}$ per $\ell^2$ per unit $\tilde t$ for $\Delta\tilde t=1$, $2$, $4$ and $8$.
Birth rates are therefore comparable only at a fixed frame interval, which is why we compare masses within the regime scan and sizes within the size scan.

\begin{table}[b]
\caption{Defect density $10^3n_d\ell^2$ at $\mt=1$ for several detector settings and system sizes. Default: $s=\ell$, threshold $0.2$, mask $0.3$, loop half-width of 3 grid spacings; each other row changes one parameter. The shuffled control gives at most $0.04$ in these units for every entry.}
\label{tab:defects}
\begin{ruledtabular}
\begin{tabular}{lcccccc}
 & \multicolumn{3}{c}{$\kt=0.5$} & \multicolumn{3}{c}{$\kt=2$}\\
$L/\ell$ & 16 & 24 & 48 & 16 & 24 & 48\\
\hline
default & 14.1 & 11.1 & 17.3 & 0.2 & 0.1 & 0.2\\
$s=0.75\ell$ & 10.1 & 5.4 & 9.4 & 12.1 & 8.5 & 8.4\\
$s=\Lambda/5$ & 10.1 & 11.2 & 14.6 & 10.1 & 7.0 & 6.9\\
threshold $0.3$ & 7.7 & 5.9 & 9.2 & 0 & 0 & 0\\
mask $0.5$ & 5.6 & 3.1 & 7.4 & 0.2 & 0.1 & 0.1\\
loop half-width 4 & 6.9 & 6.8 & 9.4 & 0.03 & 0.02 & 0.01\\
\end{tabular}
\end{ruledtabular}
\end{table}
\section{Weakly nonlinear analysis of the filament Hopf bifurcation}\label{app:hopf}
We use the rescaled units of Eq.~\eqref{eq:groups} throughout.

\emph{Reduced system.}
For a straight filament along $\hat{\mathbf{x}}$, $H$ is invariant under $y\to-y$, so the configuration with all particles on the line is an invariant subspace, and the dynamics within it is exact.
Its state is $\mathbf{z}=(x_i,\theta_i,\dot x_i,\dot\theta_i)_{i=1}^N$.
We take the ring of the simulations: length $L=2\pi n_w\ell/q$ with $n_w$ windings, $N=\rho_1L$ particles, and the truncated, shifted kernel of Sec.~\ref{sec:numerics} with line normalization.
The equilibrium is $x_i^*=(i-\tfrac12)L/N$, $\theta_i^*=qx_i^*/\ell$.
We write the equations of motion as $\dot{\mathbf{z}}=\mathbf{F}(\mathbf{z};\mt)$ and expand about the equilibrium as $\mathbf{F}=\mathsf{M}\mathbf{u}+\tfrac12\mathsf{B}(\mathbf{u},\mathbf{u})+\tfrac16\mathsf{C}(\mathbf{u},\mathbf{u},\mathbf{u})+\dots$, with $\mathbf{u}=\mathbf{z}-\mathbf{z}^*$.
The multilinear forms $\mathsf{B}$ and $\mathsf{C}$ are evaluated exactly by automatic differentiation.

\emph{Symmetries.}
Uniform translation along the line and a uniform phase shift leave $\mathbf{F}$ invariant and give two zero eigenvalues of $\mathsf{M}$.
Because the pair forces and torques are antisymmetric, $\sum_i(x_i+\mt\dot x_i)$ and $\sum_i(\theta_i+\mt\dot\theta_i)$ are conserved for every configuration.
They are therefore left null vectors of $\mathsf{M}$, and their projections of $\mathsf{B}$ and $\mathsf{C}$ vanish identically.
The shift $i\to i+1$, combined with $\theta\to\theta-qL/(N\ell)$, maps the equilibrium to itself and commutes with $\mathsf{M}$, so the eigenvectors of $\mathsf{M}$ are Bloch modes with wavenumbers $K=2\pi j/L$.
The reflection $x\to-x$, $\theta\to-\theta$ maps $K$ to $-K$, so every complex eigenvalue with $K\ne0$ is doubly degenerate.

\emph{Hopf point.}
By Eq.~\eqref{eq:quad}, a mode whose overdamped eigenvalue is $s$ reaches $\lambda=i\omega$ at $\mt=-\Re s/(\Im s)^2$ with $\omega=\Im s$, and the threshold $\mt_c$ is the minimum of this over all modes.
Differentiating Eq.~\eqref{eq:quad} at fixed $s$ gives $d\lambda/d\mt=-\lambda^2/(1+2\mt\lambda)$, hence
\begin{equation}
\left.\frac{d\Re\lambda}{d\mt}\right|_{\lambda=i\omega}=\frac{\omega^2}{1+4\mt^2\omega^2}>0 .
\label{eq:transv}
\end{equation}
Every complex mode therefore crosses the imaginary axis transversally, from stable to unstable, as the mass increases.
Table~\ref{tab:hopf} lists the critical quantities for the two filaments of Fig.~\ref{fig:thrash}; diagonalizing the full Jacobian gives the same $\mt_c$ to $10^{-5}$ and the same crossing speed to six digits.

\emph{Normal form.}
Let $\mathbf{q}_{1,2}$ be the critical eigenvectors with $K=\pm K_c$ and eigenvalue $i\omega_c$, normalized to $|\mathbf{q}_j|=1$, and let $\mathbf{p}_{1,2}$ be the left eigenvectors with $\mathbf{p}_j^{\mathsf T}\mathbf{q}_k=\delta_{jk}$.
On the center manifold, $\mathbf{u}=z_1\mathbf{q}_1+z_2\mathbf{q}_2+\mathrm{c.c.}+O(|z|^2)$, and the O(2) symmetry restricts the amplitude equations to cubic order to \cite{Golubitsky1988}
\begin{equation}
\dot z_1=z_1\left(i\omega_c+\sigma+a|z_1|^2+b|z_2|^2\right),
\label{eq:nf}
\end{equation}
together with the same equation with $z_1\leftrightarrow z_2$.
Here $\sigma\simeq(\mt-\mt_c)\,d\lambda/d\mt$, and no lattice resonance enters at cubic order for the filaments considered.
The coefficients follow from the standard center-manifold reduction \cite{Kuznetsov2004}:
\begin{align*}
a&=\mathbf{p}_1^{\mathsf T}\!\left[\tfrac12\mathsf{C}(\mathbf{q}_1,\mathbf{q}_1,\bar{\mathbf{q}}_1)+\mathsf{B}(\mathbf{q}_1,\mathbf{h}_{1\bar1})+\mathsf{B}(\bar{\mathbf{q}}_1,\mathbf{h}_{11})\right],\\
b&=\mathbf{p}_1^{\mathsf T}\big[\mathsf{C}(\mathbf{q}_1,\mathbf{q}_2,\bar{\mathbf{q}}_2)+\mathsf{B}(\mathbf{q}_1,\mathbf{h}_{2\bar2})\\
&\qquad+\mathsf{B}(\mathbf{q}_2,\mathbf{h}_{1\bar2})+\mathsf{B}(\bar{\mathbf{q}}_2,\mathbf{h}_{12})\big],
\end{align*}
with $\mathbf{h}_{11}=\tfrac12(2i\omega_c-\mathsf{M})^{-1}\mathsf{B}(\mathbf{q}_1,\mathbf{q}_1)$, $\mathbf{h}_{12}=(2i\omega_c-\mathsf{M})^{-1}\mathsf{B}(\mathbf{q}_1,\mathbf{q}_2)$, and $\mathbf{h}_{1\bar1}$, $\mathbf{h}_{1\bar2}$, $\mathbf{h}_{2\bar2}$ the solutions of $-\mathsf{M}\mathbf{h}=\mathsf{B}(\mathbf{q}_1,\bar{\mathbf{q}}_1)$, $\mathsf{B}(\mathbf{q}_1,\bar{\mathbf{q}}_2)$ and $\mathsf{B}(\mathbf{q}_2,\bar{\mathbf{q}}_2)$.
For the two zero-frequency equations with zero net wavenumber, $\mathsf{M}$ is singular; the right-hand sides are orthogonal to its left null vectors, as shown above, so the equations are solvable.
Their solutions are defined only up to the symmetry directions, which do not affect $a$ and $b$ because $\mathsf{B}$ vanishes along them.
Equation~\eqref{eq:nf} has a traveling-wave branch ($z_2=0$, $|z_1|^2=-\Re\sigma/\Re a$) and a standing-wave branch ($|z_1|=|z_2|$, $|z_1|^2=-\Re\sigma/\Re(a+b)$).
Each is supercritical when its denominator is negative, and when both are, the traveling wave is the stable one if $\Re b<\Re a$ \cite{Golubitsky1988}.

\begin{table}[b]
\caption{Hopf bifurcation of the straight filament with $q\ell=0.95$ at $\kt=0.25$ (two lengths) and $\kt=0.15$. The coefficients refer to Eq.~\eqref{eq:nf}; ratios are independent of the normalization of $\mathbf{q}_j$.}
\label{tab:hopf}
\begin{ruledtabular}
\begin{tabular}{lccc}
 & \multicolumn{2}{c}{$\kt=0.25$} & $\kt=0.15$\\
 & $N=159$ & $N=317$ & $N=159$\\
\hline
$\mt_c$ & 0.12906 & 0.12945 & 0.4717\\
$\tilde\omega_c$ & 0.3901 & 0.3889 & 0.313\\
$K_c\ell$ & 2.533 & 2.533 & 2.375\\
$d\Re\tilde\lambda/d\mt$, Eq.~\eqref{eq:transv} & 0.1506 & 0.1497 & 0.0903\\
$\Re a$ & $<0$ & $<0$ & $<0$\\
$\Im a/\Re a$ & 6.98 & 6.98 & 4.9\\
$\Re b/\Re a$ & 149.2 & 149.2 & 152\\
$N\Re a/(N\Re a)_{N=159}$ & 1 & 0.996 & --\\
\end{tabular}
\end{ruledtabular}
\end{table}

Table~\ref{tab:hopf} shows that $\Re a<0$ and $\Re(a+b)=150\,\Re a<0$ for both filaments.
Both branches are therefore supercritical, and the traveling wave is stable.
The coefficient $a$ scales as $1/N$ only because of the normalization $|\mathbf{q}_j|=1$; the intensive combination $N\Re a$ agrees between the two filaments to 0.4\%, as do the ratios, so the result does not depend on the filament length.
Expressed through the kinetic energy, the traveling-wave branch reads $\langle|\dot{\mathbf{x}}|^2\rangle\approx0.25(\mt-\mt_c)$ to cubic order.

\emph{Second parameter point.}
To check that supercriticality is not special to $\kt=0.25$, we repeated the reduction at $\kt=0.15$ (last column of Table~\ref{tab:hopf}).
The threshold moves to $\mt_c=0.4717$ and the frequency drops to $\tilde\omega_c=0.313$, but both branches are again supercritical and the traveling wave is again stable, with ratios $\Im a/\Re a$ and $\Re b/\Re a$ close to those at $\kt=0.25$.
Upward runs from a straight filament perturbed with amplitude $10^{-3}$ ($r_c=2.52\ell$, see Sec.~\ref{sec:ignite}) stay at rest at $\mt=0.40$ and $0.45$ and set into motion at $\mt=0.5$, $0.6$ and $0.8$, after delays that shorten from $\tilde t\approx1400$ to $150$ as the linear growth rate rises.
With $r_c=2.5\ell$ the runs at $\mt=0.3$ and $0.4$ also escape, after $\tilde t\approx600$--$1300$, through the cutoff-triggered pairing described in Sec.~\ref{sec:ignite}.

\begin{figure}[tb]
\includegraphics[width=\columnwidth]{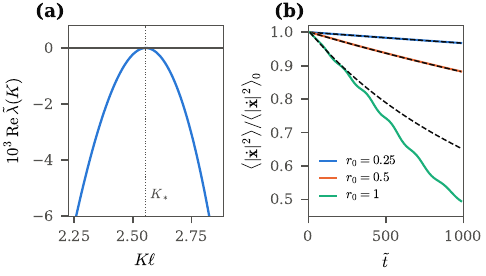}
\caption{Weakly nonlinear analysis of the filament Hopf bifurcation ($\kt=0.25$, $q\ell=0.95$).
(a) Growth rate of the critical branch against continuous wavenumber for an unbounded filament at its threshold $\mt_c^\infty=0.1292$; the maximum lies at $K_*\ell=2.555$.
(b) Decay of a traveling wave started on the critical mode at $\mt=\mt_c$ ($N=159$), for three initial amplitudes, averaged over one Hopf period (solid), compared with the cubic normal form, Eq.~\eqref{eq:nf} (dashed).}
\label{fig:hopf}
\end{figure}

\emph{Direct check.}
At $\mt=\mt_c$ the linear growth vanishes, and Eq.~\eqref{eq:nf} predicts that a traveling wave of initial amplitude $r_0$ decays as $|z_1|^2=r_0^2/(1-2\Re a\,r_0^2t)$.
We integrated the reduced equations from $\mathbf{z}^*+2\Re(r_0\mathbf{q}_1)$ with a fourth-order Runge--Kutta scheme [Fig.~\ref{fig:hopf}(b)].
For $r_0=0.25$ and $0.5$ the kinetic energy follows the prediction to within $0.1\%$ over $\tilde t=1000$.
At $r_0=1$, where $\langle|\dot{\mathbf{x}}|^2\rangle\approx4\times10^{-4}$, it decays faster than predicted, so the higher-order terms are stabilizing as well.
No unstable cycle exists at these amplitudes.

\emph{Long filaments.}
For an unbounded filament the wavenumber is continuous, and the $4\times4$ Bloch block $\mathsf{M}(K)=\sum_j\mathsf{M}_{0j}e^{iK(x_j^*-x_0^*)}$ gives the dispersion relation $\lambda(K)$ [Fig.~\ref{fig:hopf}(a)].
Its maximum reaches zero at $\mt_c^\infty=0.1292$, at $K_*\ell=2.555$, with $\tilde\omega=0.386$.
A slowly modulated wave $\mathbf{u}=A(X,t)e^{iK_*X}\mathbf{q}+\mathrm{c.c.}$ obeys the complex Ginzburg--Landau equation
\begin{equation*}
\partial_tA=\sigma A+v_g\partial_XA+d\,\partial_X^2A+a|A|^2A ,
\end{equation*}
with $d=-\tfrac12\lambda''(K_*)$ and $v_g=-i\lambda'(K_*)$.
We find $\tilde v_g=-0.135$ and $d=0.0770+0.0912i$, so $c_1=\Im d/\Re d=1.19$ and $c_3=\Im a/\Re a=6.98$.
The Benjamin--Feir--Newell criterion gives $1+c_1c_3=9.3>0$, so the uniform traveling wave is stable to long-wave modulations \cite{Aranson2002}.
This equation omits the coupling to the long-wave modes of the two conserved quantities identified above, which in a long filament are slow and can destabilize the wave \cite{Matthews2000}.
Within the cubic theory, however, the irregular moving state of Fig.~\ref{fig:thrash} is not a modulational instability of the bifurcating wave.

\section{Convergence in the number of particles per kernel}\label{app:nell}

To test how the results depend on particle discreteness we repeated the bulk runs at $\mt=0.25$ and $L=24\ell$ at $n_\ell=12.5$, $25$, $50$ and $100$, by setting $\ell=\sqrt2$, $2$, $2\sqrt2$ and $4$ at fixed $\rho=1$ ($N=1152$ to $9216$).
The physical system is the same in all four cases; only the number of particles that represent it changes.
The time step in rescaled units is the same at all resolutions, each run lasts $\tilde t=250$ with averages over the second half, and we used two initial conditions for $n_\ell\le50$ and one for $n_\ell=100$.
Defects are detected on ten snapshots per run with the adaptive coarse-graining scale $s=\Lambda/5$, which fixes the detection scale relative to the texture rather than to the particle spacing, and compared with a shuffled-phase control.
The sum rule, Eq.~\eqref{eq:sumrule}, holds to within 1\% in every run (3\% at $\kt=6$, $n_\ell=100$).

Table~\ref{tab:nell} lists the results, from which four conclusions follow.
(i) In the filament regime ($\kt\le2$) none of the observables shows a systematic trend with $n_\ell$: the differences between resolutions are comparable to the differences between initial conditions at the same resolution.
(ii) In the weak texture ($\kt=3$, $4$) $\Lambda$, $R$ and the speed are converged from $n_\ell=25$ onward, while $A$ increases by about $0.1$ between $n_\ell=25$ and $50$ and is then constant; $n_\ell=12.5$ under-resolves the texture at $\kt=4$.
(iii) For $n_\ell\ge25$ the defect density at fixed $s/\Lambda$ shows no trend with $n_\ell$ for $\kt\le3$, where $n_d\ell^2\approx(7$--$18)\times10^{-3}$, and neither do the birth rates for $\kt\le2$.
At $\kt=4$ the density decreases from $5.2\times10^{-3}$ at $n_\ell=25$ to $3.0\times10^{-3}$ at $n_\ell=100$, a change comparable to the scatter between initial conditions, so we cannot tell whether it is significant.
The shuffled control is clean for $n_\ell\ge50$; at $n_\ell=12.5$ it reaches a level comparable to the signal at $\kt=4$, so the defect counts at that resolution are unreliable.
With the fixed scale $s=\ell$, the detector finds $(1$--$2)\times10^{-3}$ defects per $\ell^2$ at $\kt=2$ and none at $\kt\ge3$ at every resolution, which confirms that its failure in the weak texture is set by the ratio $s/\Lambda$, not by the particle number.
(iv) In the phase liquid ($\kt=6$) the residual speed and $R$ decrease with $n_\ell$, roughly as $n_\ell^{-0.7}$--$n_\ell^{-0.8}$, so this motion is a finite-$n_\ell$ effect.

The averaging window is short compared with the slowest relaxation in the weak texture and the phase liquid.
The last column of Table~\ref{tab:nell} gives the relative change of $\langle|\dot{\mathbf{x}}|^2\rangle$ between the two halves of the window.
In the filament regime it lies between $+5\%$ and $+25\%$ with no trend in $n_\ell$, but at $\kt=4$ and $6$ it grows with $n_\ell$ and reaches $+60\%$ at $n_\ell=100$, so those two runs have not reached a stationary state.
The exponent quoted in (iv) is therefore based on $n_\ell\le50$, where the drift is below 10\%.

\begin{table}[tb]
\caption{Dependence on the number of particles per kernel at $\mt=0.25$, $L=24\ell$. $n_{\rm ic}$: number of initial conditions; $v$: rms speed (rescaled units); $n_d$: defects per $10^3\ell^2$ detected with $s=\Lambda/5$, with the shuffled control in parentheses; $\delta$: change of $\langle|\dot{\mathbf{x}}|^2\rangle$ between the first and second halves of the averaging window. Values are means over initial conditions.}
\label{tab:nell}
\begin{ruledtabular}
\begin{tabular}{ccccccccc}
$\kt$ & $n_\ell$ & $n_{\rm ic}$ & $v$ & $R$ & $\Lambda/\ell$ & $A$ & $n_d$ & $\delta$ (\%) \\
\hline
$0.5$ & 12.5 & 2 & 0.338 & 0.507 & 5.58 & 0.93 & 12.3 (0.0) & $+9$ \\
 & 25 & 2 & 0.356 & 0.522 & 5.81 & 0.92 & 9.1 (0.0) & $+25$ \\
 & 50 & 2 & 0.333 & 0.514 & 5.62 & 0.95 & 14.8 (0.0) & $+19$ \\
 & 100 & 1 & 0.347 & 0.526 & 5.74 & 1.00 & 10.8 (0.0) & $+6$ \\
\hline
$1$ & 12.5 & 2 & 0.401 & 0.402 & 5.04 & 0.81 & 13.4 (0.0) & $+5$ \\
 & 25 & 2 & 0.384 & 0.397 & 4.91 & 0.91 & 17.9 (0.0) & $+21$ \\
 & 50 & 2 & 0.426 & 0.410 & 5.10 & 0.92 & 12.0 (0.0) & $+11$ \\
 & 100 & 1 & 0.403 & 0.412 & 4.98 & 0.96 & 14.9 (0.0) & $+10$ \\
\hline
$2$ & 12.5 & 2 & 0.306 & 0.218 & 3.95 & 0.61 & 13.5 (0.7) & $+19$ \\
 & 25 & 2 & 0.307 & 0.218 & 3.96 & 0.69 & 11.8 (0.0) & $+9$ \\
 & 50 & 2 & 0.308 & 0.215 & 3.96 & 0.67 & 11.5 (0.0) & $+13$ \\
 & 100 & 1 & 0.290 & 0.221 & 3.97 & 0.73 & 11.3 (0.0) & $+7$ \\
\hline
$3$ & 12.5 & 2 & 0.191 & 0.121 & 3.41 & 0.47 & 12.3 (1.7) & $-3$ \\
 & 25 & 2 & 0.216 & 0.121 & 3.44 & 0.50 & 7.2 (0.4) & $+9$ \\
 & 50 & 2 & 0.203 & 0.119 & 3.40 & 0.59 & 7.6 (0.0) & $+9$ \\
 & 100 & 1 & 0.207 & 0.121 & 3.40 & 0.61 & 8.2 (0.0) & $+23$ \\
\hline
$4$ & 12.5 & 2 & 0.097 & 0.064 & 3.03 & 0.17 & 10.1 (6.3) & $+4$ \\
 & 25 & 2 & 0.128 & 0.070 & 3.08 & 0.40 & 5.2 (0.4) & $+15$ \\
 & 50 & 2 & 0.136 & 0.075 & 3.05 & 0.51 & 3.8 (0.0) & $+30$ \\
 & 100 & 1 & 0.127 & 0.069 & 3.05 & 0.48 & 3.0 (0.0) & $+64$ \\
\hline
$6$ & 12.5 & 2 & 0.057 & 0.035 & -- & 0 & -- & $-2$ \\
 & 25 & 2 & 0.032 & 0.018 & -- & 0 & -- & $+5$ \\
 & 50 & 2 & 0.019 & 0.011 & -- & 0 & -- & $+7$ \\
 & 100 & 1 & 0.014 & 0.008 & -- & 0 & -- & $+59$ \\
\end{tabular}
\end{ruledtabular}
\end{table}


\end{document}